# A comprehensive study of comet 67P/Churyumov-Gerasimenko in the 2021/2022 apparition. II. Colorimetry, polarimetry, modeling

Vera Rosenbush [a], Oleksandra Ivanova [b,c,*], Johannes Markkanen [d], Igor Lukyanyk [a], Valerii Kleshchonok [a,d], Ludmilla Kolokolova [e], Colin Snodgrass [f], Elena Shablovinskaya [g]

[a] *Astronomical Observatory of Taras Shevchenko National University of Kyiv, 3 Observatorna St., 04053 Kyiv, Ukraine*
[b] *Astronomical Institute of Slovak Academy of Sciences, 059 60 Tatranska Lomnica, Slovak Republic*
[c] *Main Astronomical Observatory of National Academy of Sciences of Ukraine, 27 Akademika Zabolotnoho St., 03143 Kyiv, Ukraine*
[d] *Institut für Geophysik und Extraterrestrische Physik, Technische Universität Braunschweig, Mendelssohnstr. 3, D-38106 Braunschweig, Germany*
[e] *University of Maryland, College Park, MD, USA*
[f] *Institute for Astronomy, University of Edinburgh, Edinburgh EH9 3HJ, UK*
[g] *Núcleo de Astronomá de la Facultad de Ingenierá, Universidad Diego Portales, Av. Ejército Libertador 441, Santiago, Chile*

This paper presents the second part of our multi-technique observational study of comet 67P/Chur-yumov–Gerasimenko during its 2021/22 apparition. While the first paper (Paper I; Rosenbush et al., 2026) focused on photometry and spectroscopy, here we report observations of dust color and linear polarization across the coma. The dataset combines imaging obtained with the *g*-sdss and *r*-sdss filters at the 6-m BTA SAO Telescope with aperture polarimetry performed at the 2.6-m Shajn telescope (CrAO) and the 2-m RCC telescope (Peak Terskol Observatory). The imaging observations were acquired at phase angles of 47.9° (6 October 2021) and 10.5° (6 February 2022). They show negative polarization of about 0.5–2.5% at 10.5° and positive polarization averaging ~11% at 47.9°. The color was red in both cases, with values of ~0.8 mag near the nucleus and in the tail, decreasing to ~0.2 mag in the coma and jet at nucleocentric distances up to 20,000 km. We constructed spatial maps of dust color and linear polarization across the coma and tail. The images reveal a generally smooth decrease in color index and a corresponding increase in polarization with increasing distance from the nucleus. Numerical modeling of the observed color and polarization maps was performed using ballistic particle–cluster aggregates (BPCA) composed of equal fractions of silicates and organics, incorporating both light-scattering and dynamical simulations. The modeling indicates that the coma is dominated, at all distances, by particles larger than 10 μm. Small-scale variations in the polarization maps likely correspond to regions enriched in smaller particles (1–10 μm). The aperture data were combined with published polarization measurements of comet 67P/ C–G from four previous apparitions. These results confirm that the comet belongs to the high-polarization class. A notable feature of the 2021/22 apparition is that the positive polarization appears higher than is typical even for high-polarization comets. This behavior, together with the relatively isotropic coma structure, may indicate changes in the dust properties that occurred after the 2015/16 apparition.

The Jupiter-family comet 67P/Churyumov–Gerasimenko (hereafter 67P/C–G) is one of the best- studied comets, primarily owing to the Rosetta mission and extensive coordinated ground-based observing campaigns (see, for example, one of them Snodgrass et al., 2017, and numerous chapters in Thomas et al., 2021). In addition to the in-situ and remote-sensing data obtained by Rosetta, our group conducted coordinated quasi-simultaneous photometric, spectroscopic, and polarimetric observations during the 2015/16 apparition (Rosenbush et al., 2017; Ivanova et al., 2017a), providing a consistent multi-technique characterization of the comet's dust and gas environment around perihelion. A similar observing strategy was applied during the 2021/22 apparition of 67P/C–G, combining nearly simultaneous spectroscopic and photometric monitoring with polarimetric measurements.

In our previous paper (Rosenbush et al., 2026, Paper I hereafter),

which is the first part of our research of comet 67P/C–G, we presented the results of photometric and spectroscopic observations of the comet obtained with the 6-m BTA SAO telescope and the 2-m Liverpool Telescope. The comet showed a high continuum level in its spectrum, a fairly strong CN emission, and relatively weak $C_2$, $C_3$, and $NH_2$ emissions. Comet 67P/C–G has been confirmed to be a $C_2$-depleted comet, which is consistent with previous studies (e.g., Schulz et al., 2004; Ivanova et al., 2017a). On average, the normalized reflectivity gradient of the dust was $\sim$ 16%/1000 on 6 October 2021 and about 14%/1000 on 6 February 2022 within the wavelength range of 4450–6839 . The photometric observations revealed pronounced temporal and morphological changes in the dust coma across perihelion (2 November 2021 at $r = 1.211$ au), including the development of sunward jets, a neckline structure, and an unusual CN coma morphology. Dust activity decreased significantly: $Af\rho$ in the *r*-sdss filter dropped from 943 $\pm$ 14 cm on 6 October 2021 to 286 $\pm$ 7 cm on 6 February 2022. During both observing periods, the surface-brightness profiles of the tail differed significantly from those of the coma and jets. Notably, the comet exhibited higher dust production than during the 2015/16 apparition (Rosenbush et al., 2017).

The enhanced images of comet 67P/C–G obtained pre-perihelion show a jet-like structure directed toward the Sun at the position angle $PA_{\rm jet} \approx 127^\circ$ and a dust tail in the antisolar direction at $PA_{\rm tail} \approx 266^\circ$. Using a geometric model (Kleshchonok and Sierks, 2025) to simulate the jet formation process, the latitude of the active area on the nucleus, $\varphi = -70^\circ \pm 4^\circ$, the jet opening angle of $20^\circ \pm 6^\circ$, and the particle ejection velocity of 0.30 $\pm$ 0.04 km/s were determined. After perihelion, on 6 February 2022, the structure of the dust coma changed; a neckline structure appeared in the solar direction ($PA_{\rm NL} = 284^\circ$), along with two jet-like structures, one of which is directed perpendicular to the solar-antisolar direction at $PA_{\rm J1} = 193^\circ$, and the second toward the antisolar direction at $PA_{\rm J2} = 133^\circ$. According to modeling, both jets originate from the same active source on the rotating nucleus located at a latitude of $\varphi = -58^\circ \pm 5^\circ$ and having a jet opening angle of $26^\circ \pm 8^\circ$. Moreover, there is another active source located at latitude $\varphi = -53^\circ \pm 10^\circ$ and separated from the first one by approximately $150^\circ \pm 20^\circ$ in longitude. This active area generates the third jet. The first active region may be located on a convex part of the nucleus surface near the border between the Khonsu and Bes regions, while the second region may be located in a depression near the border between Anhur and Bes, receiving less sunlight than the first region. The mean particle velocity in the jets is estimated to be about 0.32 $\pm$ 0.04 km/s.

Building on these results, the present work focuses on the analysis of dust color and polarization maps.

of the coma, together with numerical modeling of the dust environment. This combined approach provides deeper insight into the physical properties of the dust grains, including their size, composition, and structure, and allows us to investigate how these parameters evolve throughout the 2021/22 apparition.

This paper is organized as follows. Section 2 describes the imaging photometric and polarimetric observations of comet 67P/C–G, as well as the aperture polarization measurements, and data reduction. Section 3 presents the spatial distribution of color, and Section 4 reports the results of the linear polarimetry. Numerical modeling of the color and polarization variations across the coma is described in Section 5. In Section 6, we compare and discuss the 2015/16 and 2021/22 apparitions, and Section 7 summarizes the conclusions.

## 2. Observations and data processing

### 2.1. Overview of observations

In this work, we analyze photometric and polarimetric observations of comet 67P/C–G obtained at three telescopes: the 6-m ($f$/4) BTA (Big *Telescope* Alt-Azimuthal) telescope (SAO), the 2.6-m ($f$/16) Shajn telescope of the Crimean Astrophysical Observatory (CrAO) and the 2-m ($f$/8) Ritchey–Chrétien–Coudé *telescope* of the International Center for Astronomical, Medical and Ecological Research (Peak Terskol Observatory).

The main dataset was acquired with the 6-m BTA telescope using the multi-mode focal reducer SCORPIO-2 (Afanasiev and Moiseev, 2011; Afanasiev and Amirkhanyan, 2012). To construct the polarization phase curve, additional aperture polarimetric observations were obtained using the photoelectric polarimeters at the CrAO and Terskol Observatory (Kiselev et al., 2022).

In Table 1 we provide the geometric and technical details of the observations of comet 67P/C–G performed with the 6-m BTA (SAO), 2.6-m Shajn (CrAO) and the 2-m RCC (Peak Terskol Observatory) telescopes, including the observation time (mid-cycle, UT), heliocentric ($r$) and geocentric ($\Delta$) distances, the phase angle of the comet ($\alpha$), the position angle ($\varphi$) of the scattering plane, the filter, the aperture radius in arcseconds and kilometers, the total exposure time ($T_{\rm exp}$), the number of observation cycles ($N$), and the observation mode, the used telescope, and the observatory.

### 2.2. Observations at the 6-m BTA telescope

#### 2.2.1. Instrumentation and observing runs

Photometric and polarimetric observations of comet 67P/C–G were performed using the multi-mode focal reducer SCORPIO-2 (Spectral Camera with Optical Reducer for Photometrical and Interferometrical Observations), mounted at the primary focus of the telescope. This instrument was equipped with an E2V CCD261–84 detector (2048 $\times$ 4104 pixels, 15 μm), providing an image scale of 0.4 arcsec/pixel and a field of view of $6.8' \times 6.8'$. A 2 $\times$ 2 on-chip binning was applied to improve the signal-to-noise ratio.

The comet was observed at two epochs:

- 6 October 2021 (31 days before perihelion; $r = 1.254$ au, $\Delta = 0.475$ au, phase angle $\alpha = 47.9^\circ$);
- 6 February 2022 (96 days after perihelion; $r = 1.675$ au, $\Delta = 0.709$ au, phase angle $\alpha = 10.5^\circ$).

The atmospheric seeing was $\sim 2''$ ($\sim$ 690 km at the comet) in October and $\sim 3''$ ($\sim$ 1030 km) in February. Guiding was performed at the comet's apparent motion. Flat-field calibration was obtained using twilight sky exposures.

The standard procedure of reductions included bias subtraction, flat-field correction, cosmic ray removal, frame alignment, and robust stacking. Final combined images were produced for each observing night in both photometric and polarimetric modes. A detailed description of the data reduction, sky background treatment, and error analysis is given in Paper I.

#### 2.2.2. Photometry

Photometric observations on 6 October 2021 were carried out using the SDSS (Sloan Digital Sky Survey) broadband filters *g*-sdss (the central wavelength and Full Width at Half Maximum (FWHM) are 4650/1300 Å) and *r*-sdss (6200/1200 Å). For each frame, the sky background was estimated in regions free of the coma, tail, and field stars. Calibration of the photometric data was performed using field stars present in the observed frames. Stellar magnitudes for the reference stars were taken from the APASS catalogue (AAVSO Photometric All-Sky Survey), with typical uncertainties of $\sim$0.02 mag, depending on the star's brightness.

Initial image reduction followed standard procedures, including bias subtraction, flat-field correction, and removal of cosmic rays. The calibrated photometric data were then used to construct color maps of the comet, allowing us to investigate variations in dust properties across the coma and tail. Despite the fact that on 6 February 2022 we observed the comet with narrowband comet HB filters (Farnham et al., 2000), namely the continuum *BC* (4450/62 Å) and *RC* (6839/96 Å) filters designed to isolate the continuum, the color analysis was performed using broadband filter images. Since standards for the narrowband filters had not been obtained, there was no corresponding narrowband standard system.

**Table 1**
Log of observations of comet 67P/Churyumov–Gerasimenko at the 6-m BTA (SAO), 2.6-m Shajn (CrAO), and the 2-m RCC (Peak Terskol Observatory) telescopes.

| Date, UT | $r$ [au] | Δ [au] | $\alpha$ [deg] | $\varphi$ [deg] | Filter | Dia [arcsec] | Dia [km] | $T_{exp}$ [s] | $N$ | Mode | Tel [m] | Obs |
|---|---|---|---|---|---|---|---|---|---|---|---|---|
| 2021 | | | | | | | | | | | | |
| Aug 18.825 | 1.517 | 0.846 | 39.1 | 252.7 | R | 15 | 9204 | – | – | AperPol | 2.6 | CrAO |
| Aug 18.836 | 1.517 | 0.846 | 39.1 | 252.7 | *V* | 15 | 9204 | – | – | AperPol | 2.6 | CrAO |
| Aug 18.958 | 1.517 | 0.846 | 39.1 | 252.7 | *I* | 15 | 9204 | – | – | AperPol | 2.6 | CrAO |
| Sep 09.897 | 1.378 | 0.634 | 42.6 | 252.7 | *R* | 15 | 6897 | – | – | AperPol | 2.6 | CrAO |
| Sep 09.915 | 1.378 | 0.634 | 42.6 | 252.7 | *V* | 15 | 6897 | – | – | AperPol | 2.6 | CrAO |
| Sep 09.935 | 1.378 | 0.634 | 42.6 | 252. 7 | *I* | 15 | 6897 | – | – | AperPol | 2.6 | CrAO |
| Oct 6.954 | 1.254 | 0.475 | 47.9 | 266.7 | *g*-sdss | – | – | 20 | 5 | Ima | 6.0 | SAO |
| Oct 6.956 | 1.254 | 0.475 | 47.9 | 266.7 | *r*-sdss | – | – | 10 | 5 | Ima | 6.0 | SAO |
| Oct 6.959 | 1.254 | 0.475 | 47.9 | 266.7 | *R* | – | – | 30 | 16 | ImaPol | 6.0 | SAO |
| Oct 08.963 | 1.248 | 0.468 | 48.2 | 267.5 | *R* | 15 | 5091 | – | – | AperPol | 2.0 | Terskol |
| Oct 08.964 | 1.248 | 0.468 | 48.2 | 267.5 | *V* | 15 | 5091 | – | – | AperPol | 2.0 | Terskol |
| Oct 08.979 | 1.248 | 0.468 | 48.2 | 267.5 | *BC* | 15 | 5091 | – | – | AperPol | 2.0 | Terskol |
| Oct 08.980 | 1.248 | 0.468 | 48.2 | 267.5 | *RC* | 15 | 5091 | – | – | AperPol | 2.0 | Terskol |
| 2022 | | | | | | | | | | | | |
| Feb 6.897 | 1.675 | 0.709 | 10.5 | 133.5 | *r*-sdss | – | – | 15 | 15 | ImaPol | 6.0 | SAO |
| Feb 27.781 | 1.836 | 0.950 | 19.2 | 113.0 | *V* | 30 | 20,670 | – | – | AperPol | 2.0 | Terskol |
| Feb 27.781 | 1.836 | 0.950 | 19.2 | 113.0 | *R* | 30 | 20,670 | – | – | AperPol | 2.0 | Terskol |
| Feb 27.808 | 1.836 | 0.950 | 19.3 | 113.0 | *B* | 30 | 20,670 | – | – | AperPol | 2.0 | Terskol |
| Feb 27.810 | 1.836 | 0.950 | 19.3 | 113.0 | *I* | 30 | 20,670 | – | – | AperPol | 2.0 | Terskol |

The gas contribution to the broadband filters was estimated from spectroscopic data obtained during the same observing periods, as described in Paper I. We convolved the measured comet spectra with the transmission curves of the *g*-sdss and *r*-sdss filter passbands to calculate the ratio of the flux from gas emission bands (such as $C_2$ and $NH_2$) to the total flux (emission bands plus dust continuum) within each filter passband. These calculations yielded gas contributions of 7.1% and 3.6% in 2021, and 2.1% and 1.3% in 2022 for the *g*-sdss and *r*-sdss filters, respectively. Similar to the 2015/16 apparition, the gas contribution did not exceed 10%, enabling the study of the comet's dust component using broadband images.

### 2.2.3. *Imaging polarimetry*

Polarimetric observations of comet 67P/C–G at the 6-m telescope were performed with SCORPIO-2 using a dichroic polarization analyzer (POLAROID). Measurements of the linear polarization were obtained in the broadband filters *R* (6420/790 Å) on 6 October 2021 and in *r*-sdss (6200/1200 Å) on 6 February 2022. The analyzer was set at three position angles (−60°, 0°, +60°), forming a complete measurement cycle. Instrumental polarization, determined from observations of unpolarized standard stars (Serkowski, 1974), is less than 0.1% and is taken into account in the final data. The correction for the zero point of the instrumental position angle was derived from observations of highly polarized standard stars (Hsu and Breger, 1982).

A detailed description of SCORPIO-2, the observational approach with this instrument, primary processing of raw images, data reduction, calculation of polarization parameters, and error analysis can be found in Afanasiev and Moiseev (2011), Afanasiev and Amirkhanyan (2012), and Afanasiev et al. (2014). Applications of this methodology to comet observations with SCORPIO-2 in spectral, photometric, and polarimetric (linear and circular) modes are given in the previous studies: e.g., comet C/2009 P1 (Garradd) (Ivanova et al., 2017b); comet 2P/Encke (Kiselev et al., 2020; Rosenbush et al., 2020); comet 67P/Churyumov–Gerasimenko in the 2015/16 apparition (Ivanova et al., 2017a; Rosenbush et al., 2017).

## 2.3. *Aperture polarimetry*

Photoelectric polarimetric observations of comet 67P/C–G were carried out at:

– the 2.6-m Shajn telescope (CrAO) on 18 August and 9 September 2021;
– the 2-m RCC telescope (Peak Terskol Observatory) on 8 October 2021 and 27 February 2022.

Both telescopes are equipped with identically designed two-channel photoelectric polarimeters 'POLSHAKH'. The main part of the polarimeters is a modulator, which is a rapidly rotating (~30 rotations per second) achromatic waveplate with a field lens. This module is built around a hollow-shaft brushless direct-current (DC) motor with an embedded angular encoder. There are two waveplates per instrument used in the modulators: a half-wave plate (λ/2 plate) for the measurement of linear polarization and a quarter-wave plate (λ/4 plate), which is used for simultaneous measurements of linear and circular polarization. After the modulator, a Wollaston prism splits the incoming light into two orthogonally polarized beams (the ordinary, O, and extraordinary, E, beams), which form independent single-beam photo-polarimetric channels (red and blue). Thus, each channel can have its own set of spectral filters and detectors optimized for the required wavelength. The rotation of the waveplate is synchronized with the counting of photomultiplier tube (PMT) pulses in each of the 16 sectors of the waveplate position (0° − 22.5°, 22.5° − 45°, …, 337.5° − 360°) for the two orthogonally polarized beams.

This technique is based on the principle of synchronous detection, which ensures quasi-simultaneous measurements of the Stokes parameters $q$ and $u$ of the incident light, making the polarization measurements practically independent of variations in incoming intensity caused by changes in the comet's brightness or atmospheric extinction, and providing high accuracy.

The cooled Hamamatsu R943–02 PMT is used as the detector in the red channel, observing the comet in the *R* and *I* bands. For polarization measurements, the *R* filter (6880/2310 Å) is used in the CrAO polarimeter, and 6830/1590 Å in the Terskol polarimeter. The blue channel uses an uncooled EMI 6556B photomultiplier, providing observations in the *U* (3600/740 Å), *B* (4340/1170 Å), and *V* (5400/800 Å) bands. The *I* (8090/1880 Å) filter is the same in both polarimeters. Polarization of the comet was also measured in the narrowband continuum filters *BC* (4330/40 Å) and *RC* (6840/90 Å).

Polarization measurements were carried out in the following sequence: sky background – comet – sky background. The *sky background* was measured in regions close to the comet but free of the coma, tail, and background stars, and then subtracted. Instrumental polarization was determined from observations of unpolarized standard stars selected from Serkowski (1974). To calibrate the zero points of the instrumental position angles, highly polarized standard stars from Hsu and Breger (1982) and Schmidt et al. (1992) were observed. The instrumental polarization of both telescopes was less than 0.03% in all filters and was subtracted from the measured comet polarization. During the observing runs, the zero points of the polarization plane remained stable within ~1° for both telescopes. A more detailed description of the instrument and data reduction is provided by Kiselev et al. (2022).

For observations, circular diaphragms with a diameter of 15″ or 30″ were used. In Table 1, the aperture size (radius) in given in arcseconds and kilometers. For most observing nights, atmospheric seeing at the 2.6-m CrAO telescope and the 2-m telescope of the Terskol Peak Observatory was 2–3″.

## 3. Opticalcolorofthecomet

### 3.1. Color map

With the intensity images of comet 67P/C–G in the *g*-sdss and *r*-sdss filters, we derived a map of the dust color (*g-r*) following the methodology of Ivanova et al. (2021, 2023). This procedure involved several important steps: precise sky background subtraction, removal of background stars from the coma region to avoid artifacts, and sub-pixel alignment of frames using the isophote method to determine the optical center. Stars within the cometary coma were removed from the frames to prevent the formation of artifacts. The accuracy of aligning the photocenter of the color frames, as well as that of the polarization frames, was verified using the central maximum, the isophote method, and aligning images with sub-pixel accuracy method (see Paper I) to avoid potential artifacts. As in the 2015/16 observations, a bright optocenter is present in the comet, simplifying image stacking.

The color map was then constructed by calculating the color index for each pixel. The color map obtained from the observations on 6 October 2021 is shown in Fig. 1 (left panel). To highlight morphological structures with low contrast, additional digital filtering (simple rotational-gradient and asymmetry digital filters) was applied to the resulting map (see Fig. 2).

As we noted in Section 2.1, the contribution of gas emissions to the *g*-sdss and *r*-sdss filters is approximately 7.1% and 3.6% in 2021 and 2.1% and 1.3% in 2022, respectively. A simple error propagation analysis shows that the contribution from gas emissions leads to a color index change of only ~ 0.04 mag in 2021 and < 0.008 mag in 2022. This indicates that the color index is not significantly affected by gas emissions, as these changes are much smaller than the observed spatial color variations (0.2–0.8 mag) across the coma. Thus, the observed variations in color (and also polarization) maps reflect the intrinsic properties of the dust component, and the influence of gas emissions on the color index remains within the measurement errors.

As shown in the color map obtained on 6 October 2021 (Fig. 1), the dust color becomes bluer with increasing distance from the nucleus, indicating evolution of the dust particles. The right panel shows radial cuts across the coma from the optocenter in various directions. To construct these cuts, color was measured within coma areas of 3 × 3 px$^2$. The redder color near the optical center is likely due to blurring of the nucleus region caused by seeing. The radial cross-sections along the tail were extracted in the direction $PA = 266°$ (black line), along the jet at $PA = 127°$ (green line), and across the coma perpendicular to the solar-antisolar direction at $PA = 176°$ and $PA = 356°$, respectively. A radial cross-section along the tail shows no noticeable decrease in dust redness, the color remains nearly constant (on average about 0.75 mag) from the nucleus to a cometocentric distance of ~35,000 km. In contrast, cross-sections through the coma and jet show strong bluing of the dust, from ~0.85 mag near the nucleus to ~0.2 mag at cometocentric distances of about 20,000 km, reflecting some heterogeneity in the composition and size distribution of dust particles in the comet's coma.

### 3.2. Morphological features of the color map

To investigate low-contrast morphological features and reveal structures in the color distribution, we analyzed the color map using several digital filters. The map itself exhibits a high level of noise because it is derived by subtracting two images of similar intensity, requiring the use of digital filters capable of operating under elevated noise conditions. Since color maps are constructed from ratios of images obtained in different photometric bands, their signal-to-noise ratio is substantially lower than that of the original intensity images. Therefore, digital enhancement was applied only to facilitate the visualization of possible morphological structures, and not for quantitative measurements.

Fig. 2 shows the color map after applying these filters. Panel (a)

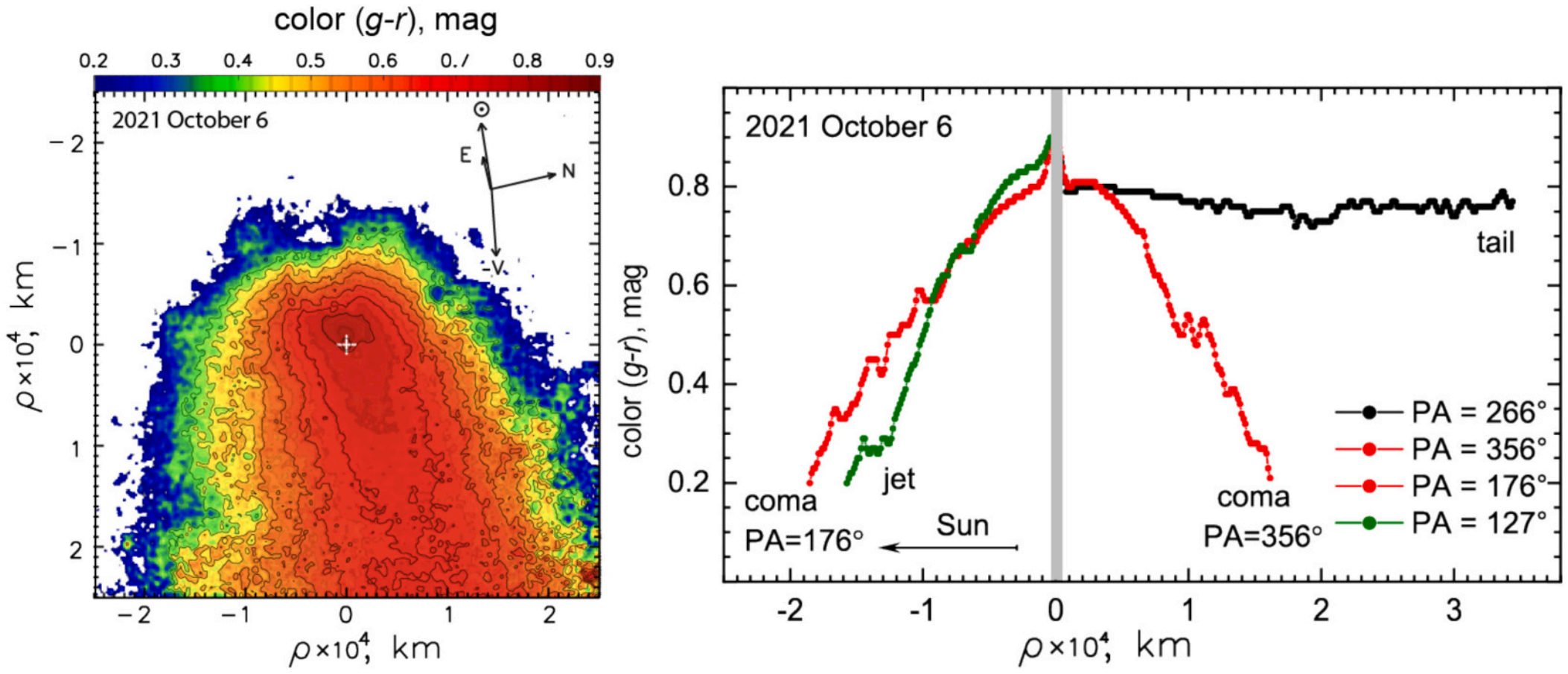

**Fig. 1.** The left panel shows the (*g–r*) color map of comet 67P/Churyumov–Gerasimenko acquired on 6 October 2021 with superimposed isophotes differing by a factor of $\sqrt{2}$ in color index. The map is colored according to the (*g-r*) color index in magnitudes as indicated by the bars at the top of the image. Additionally, the image is color-enhanced to increase contrast. In the right panel, the black line represents the cross-section along the tail in the direction $PA = 266°$, while the red lines show radial cross-cuts across the coma perpendicular to the solar-antisolar direction, in the directions $PA = 176°$and $PA = 356°$, respectively. The green line corresponds to a scan along the jet in the direction $PA = 127°$. The vertical gray strip indicates the size of the seeing disc during the observations. The zero point is at the optical center of the comet, with negative and positive distances indicating sunward and anti-sunward directions, respectively. The arrows mark the projected directions of the Sun (⊙), north (N), east (E), and the negative velocity vector of the comet projected onto the sky plane (—V). (For interpretation of the references to color in this figure legend, the reader is referred to the web version of this article.)

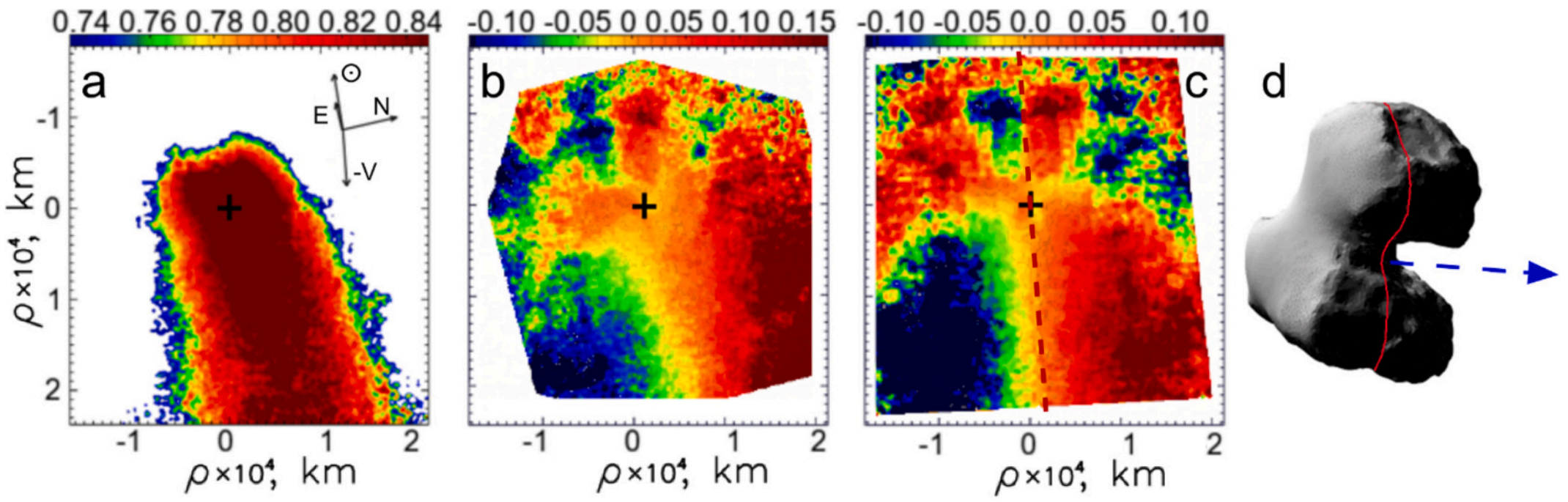

**Fig. 2.** Panel (a): Color map highlighting areas with the highest color index values. The color bar at the top of the image shows the color index in magnitudes. (b): Color map after applying a simple rotational-gradient digital filter. The color bar at the top indicates local increases or decreases in the color index in magnitudes. (c): Color map after applying the asymmetry digital filter, with the dashed line indicating the axis relative to which the filtered image was computed. The color bars at the top show the differences in color index for points located symmetrically along the filter axis. (d): View of the comet's nucleus from Earth, with the dashed line showing the orientation of the rotation axis measured from the positive (north) pole, and the red line shows the position of the equator on the nucleus. (For interpretation of the references to color in this figure legend, the reader is referred to the web version of this article.)

reproduces the image from Fig. 1 (left), but with a narrower range of color indices to emphasize the reddest regions. The commonly used rotation-gradient filter (Larson and Sekanina, 1984) is ineffective at high noise levels. Instead, we applied a simple rotational-gradient filter developed by Kleshchonok and Sierks (2025), which is more robust under such conditions. However, this filter may shift structures in position angle by up to the rotation angle used in its computation, and this should be taken into account. Panel (b) shows the result of applying the simple rotational-gradient filter with a rotation angle of 15°. The color bar at the top of the image indicates local variations in the color index in magnitudes. This filtered image clearly shows an enhancement of the color index on the northern side of the coma.

To detect deviations from symmetry with respect to the chosen axis, we applied the asymmetry filter of Ivanova et al. (2023). This filter calculates the difference between the original image and the same image rotated 180° around a selected axis, thereby revealing asymmetries in the surface brightness distribution or, in our case, in the color index. The result is shown in panel (c), where the dashed line marks the selected axis, aligned with the —V direction approximately along the comet's tail. The color scale shows the difference in magnitude between points located symmetrically relative to the axis. The filtered map shows an enhancement of the redder colors in the northern part of the coma, with a difference of approximately 0.05–0.10 mag. This enhancement is visible across nearly the entire coma, except for the southeastern region, which contains three faint background stars that distort the color in this area. These stars also produce an opposite-sign artifact in the mirror northeast region. Panel (d) shows the nucleus as seen from Earth on the observation date, with the blue dashed line indicating the orientation of the rotation axis relative to the north pole. For this geometry, the chosen asymmetry axis differs only slightly from the nucleus equator, which is shown on the nucleus by the red line. The combined analysis shows that the dust emitted from the northern and southern hemispheres differs in color. Possible explanations for this hemispheric asymmetry are discussed in the Discussion section.

We used the coordinates of the nucleus spin axis from the paper (Sierks et al., 2015), which specifies the axis coordinates as $\alpha = 69.3°$, $\delta = 64.1°$. Taking into account the comet's position and solar illumination conditions on the observation date, a 3D model was used to represent the nucleus as it would appear to an Earth-based observer. The rotation phase of the nucleus was chosen to provide the clearest visual representation. The resulting nucleus image was then rotated so that the N and E directions corresponded to the axes used in the color map (Fig. 2d). Analysis of these data suggests that the properties of the dust emitted from the northern and southern hemispheres of the nucleus differ. The simultaneous activity of the northern and southern hemispheres of the nucleus, despite the strong seasonal effects, was noted by Boehnhardt et al. (2024). Observations also show that dust can be transported even from the shadowed side of the nucleus due to the release of highly volatile species such as CO and $CO_2$ (Hässig et al., 2015).

Overall, the color map of comet 67P/C–G obtained on 6 October 2021 from the *g*-sdss and *r*-sdss images reproduces the general color distribution observed during the previous apparition on 8 November 2015 (Rosenbush et al., 2017), but with notable differences. In both apparitions, digital enhancement techniques were applied to the color maps to improve the visibility of low-contrast structures. However, while the jets are clearly visible in the 2015 color map obtained using a rotational-gradient filter, they could not be identified in the 2021 data. The jets were not visible even after applying several filtering techniques (e.g., rotational-gradient and Laplacian), including the simple rotational-gradient filter specifically designed to operate under high-noise conditions (Kleshchonok and Sierks, 2025). The radial color behavior also differs: in 2015, the jets showed a slight decrease in color index with increasing cometocentric distance relative to the surrounding coma, while in 2021, this decrease occurs in the jet somewhat faster than in the coma.

Another difference is observed in the tail color profile. In 2015, the color index decreased noticeably with cometocentric distance, whereas in 2021 the profile remains almost constant up to $\sim 3.5 \times 10^4$ km (Fig. 1, right panel). This is consistent with panel (a) (Fig. 2), which highlights regions with the highest color index and shows that the reddest particles are concentrated in the tail. In directions other than the tail, the coma exhibits a decrease in color index with increasing distance from the nucleus.

## 4. Results from polarimetry

### *4.1. Imaging polarimetry*

The polarimetric maps of comet 67P/C–G were constructed using observations obtained on 6.959 October 2021 at a phase angle of $\alpha = 47.9°$ and on 6.897 February 2022 at $\alpha = 10.54°$, using the *R* and *r*-sdss filters, respectively (Fig. 3). Sky background subtraction was carefully performed for the polarization measurements. As in our previous observations, the histogram method was employed to accurately determine and account for the sky background. Given potential weather variations during the polarimetric observations, the sky background was analyzed individually for each frame. The background was extracted from regions free of cometary coma and stars. Considering the low contribution of gas to the broadband filters, as determined from our spectral observations, the observed changes in the coma can be attributed primarily to variations in the properties of the dust. As shown in Fig. 3, the spatial distribution of polarization across the coma differs significantly between the two epochs. This difference in polarization is likely the result of a combination of phase-angle dependence (on 6 October 2021 the comet was observed 31 days before perihelion ($\alpha = 47.9^{\circ}$), whereas on 6 February 2022 it was observed 96 days after perihelion ($\alpha = 10.54^{\circ}$)) and the intrinsic evolution of the dust environment associated with the comet's orbital position (pre- and post-perihelion). Given the different observing geometries and activity states, these effects cannot be separated within the available data set. However, the consistent smoothing of the polarization at a later epoch suggests a dominant contribution from changes in dust properties.

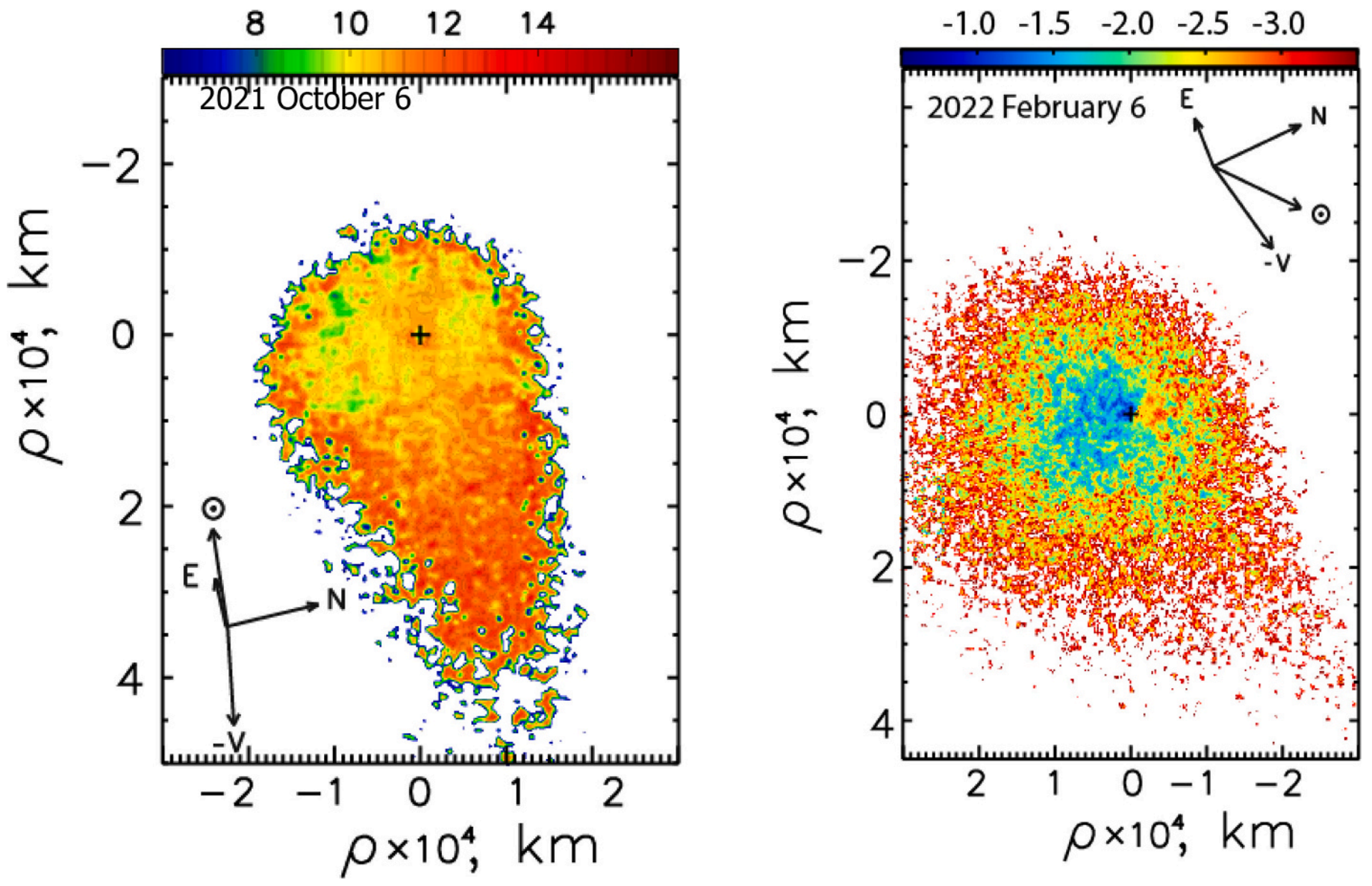

**Fig. 3.** Linear polarization maps of comet 67P/Churyumov–Gerasimenko obtained with the 6-m telescope. The left panel shows the spatial distribution of the polarization degree over the comet on 6 October 2021 in the *r*-sdss filter, while the right panel shows the observations from 6 February 2022 in the *R* filter. The color images represent the polarization values, with the corresponding scale bars in percent shown at the top of each image. The cross and arrows are the same as those shown and described in Fig. 1.

To investigate the spatial distribution of polarization across the coma of comet 67P/C–G, we measured the degree of polarization along structures identified in the brightness maps (see Figs. 3 and 4 in Paper I), namely along the jets, the tail, and in the coma perpendicular to the solar-antisolar direction. The color was also measured along the same directions (see Fig. 1). The radial profiles of polarization for the two epochs, 6 October 2021 (left panel) and 6 February 2022 (right panel), are presented in Fig. 4. Considering the scatter of data points along the cuts, it can be inferred that the spatial distribution of polarization across the coma and tail was nearly homogeneous on 6 October 2021, with values varying between 10 and 12% at a phase angle of 47.9°. The polarization in the jet, however, was systematically about 1% lower than in the surrounding coma.

It should be noted that the observational uncertainties in the radial profiles (Fig. 4) increase significantly with cometocentric distance $\rho$ due to the decreasing surface brightness of the coma and the resulting drop in the signal-to-noise ratio. Although the accuracy of aperture-averaged measurements remains high (e.g., $\sigma_P \approx 0.11\%$ for the $R$ filter on 6 October 2021, see Table 2), the point-to-point uncertainties in the radial profiles correspondingly increase in the outer regions of the coma. For a more detailed analysis of the flux error dependence on the projected distance from the optocenter and the observing conditions, we refer to Paper I.

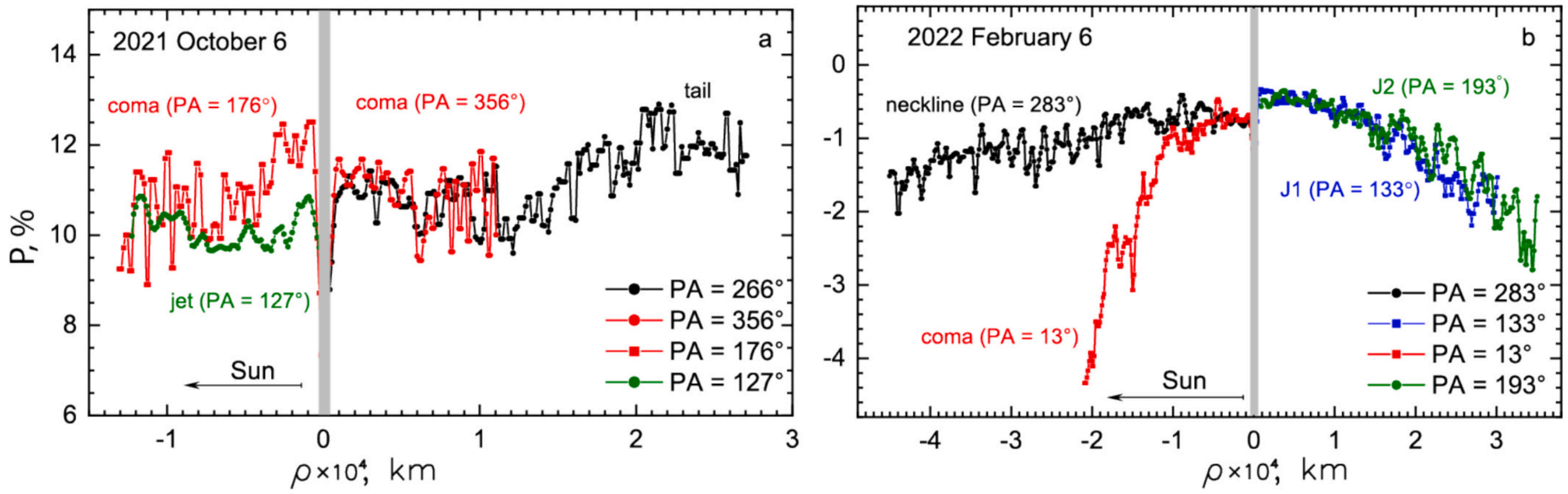

**Fig. 4.** Distribution of polarization across the coma of comet 67P/Churyumov–Gerasimenko for 31 days before perihelion (6 October 2021, left panel) and 96 days after perihelion (6 February 2022, right panel). The individual curves represent scans measured from the optocenter of the comet in different directions. In panel (a), the black line corresponds to the cut along the tail in the direction $PA = 266^{\circ}$, the red lines are the cross-cuts through the coma perpendicular to the solar-antisolar direction, in the directions $PA = 176^{\circ}$and $PA = 356^{\circ}$, respectively. The green line follows along the jet in the direction $PA = 127^{\circ}$. Panel (b) shows the cuts along the following directions: the black line is along the neckline in the sunward direction at $PA = 283^{\circ}$; the red line passes through the coma at $PA = 13^{\circ}$, opposite the bright jet at $PA = 193^{\circ}$, shown by the green line; and the blue line indicates the jet in the anti-solar direction at $PA = 133^{\circ}$. For display purposes and to illustrate the spatial trends, error bars are omitted; as detailed in Paper I uncertainty in the observations increases with increasing cometocentric distance due to the decreasing signal-to-noise ratio. Vertical gray stripes indicate the size of the seeing disk during the observations. Negative distances correspond to the solar direction, and positive distances to the anti-solar direction. (For interpretation of the references to color in this figure legend, the reader is referred to the web version of this article.)

On 6 February 2022, at a phase angle of $\alpha = 10.5^{\circ}$, the comet exhibits a more complex coma structure in polarized light. All polarization profiles show a nearly flat distribution in the near-nucleus region, with a radius of about 1000 km, where variations do not exceed ~0.5%. The degree of polarization along the neckline (see Figs. 4 and 10 in Paper I) decreases with cometocentric distance from − 1% near the nucleus to −2% at ~45,000 km. A similar trend is observed along the jets J1 and J2, where the polarization (in absolute value) increases from ~0.5% near the nucleus to ~2.5% at ~35,000 km. In the direction directly opposite the bright jet J1, the polarization decreases sharply from ~1% near the nucleus to ~4% at ~20,000 km.

Comparing our current results with previous observations at similar phase angles reveals significant changes in the polarization behavior of the coma. In the earlier post-perihelion epoch (2015/16), the polarization showed a strong spatial variability, with pronounced radial gradients and a structured negative branch (down to − 3 to − 4%) as well as large variations in positive polarization (from ~ 2% up to ~ 8%) (Rosenbush et al., 2017; Snodgrass et al., 2017). This indicates a highly inhomogeneous dust environment. In contrast, in the present observations (2021/22), the polarization field is considerably more uniform. At both large ($\alpha \approx 47.9^{\circ}$) and small ($\alpha \approx 10.5^{\circ}$) phase angles, the polarization shows weaker spatial gradients, with typical values of ~ 10–12% in the positive branch and approximately − 1 to − 2% in the negative branch, and only minor local deviations associated with variations in cometary activity. The greater global homogeneity of the coma during the 2021/22 apparition suggests changes in the comet's activity and/or dust properties.

**Table 2**
Results of photoelectric polarimetric observations of comet 67P/Churyumov–Gerasimenko acquired in the 2021/22 apparition.

| Date, UT | Filter | $\alpha$ [deg] | $\varphi$ [deg] | Dia [km] | $P \pm \sigma_P$ [%] | $\theta \pm \sigma_\theta$ [deg] | Tel [m] |
|---|---|---|---|---|---|---|---|
| 2021 Aug 18.825 | R | 39.1 | 252.7 | 9204 | 7.44 ± 0.15 | 163.84 ± 0.59 | 2.6 |
| 2021 Aug 18.836 | *V* | 39.1 | 252.7 | 9204 | 6.61 ± 0.30 | 167.45 ± 1.33 | 2.6 |
| 2021 Aug 18.958 | *I* | 39.1 | 252.7 | 9204 | 6.52 ± 0.20 | 160.34 ± 0.88 | 2.6 |
| 2021 Sep 09.897 | *R* | 42.0 | 252.7 | 6897 | 8.60 ± 0.06 | 163.99 ± 0.19 | 2.6 |
| 2021 Sep 09.915 | *V* | 42.6 | 252.7 | 6897 | 8.62 ± 0.11 | 168.77 ± 0.36 | 2.6 |
| 2021 Sep 09.935 | *I* | 42.6 | 252.7 | 6897 | 8.98 ± 0.07 | 165.04 ± 0.23 | 2.6 |
| 2021 Oct 6.959 | *R* | 47.9 | 266.7 | 4960 | 10.87 ± 0.11 | 175.42 ± 0.27 | 6.0 |
| 2021 Oct 08.963 | *R* | 48.2 | 267.5 | 5091 | 11.28 ± 0.14 | 178.53 ± 0.35 | 2.0 |
| 2021 Oct 08.964 | *V* | 48.2 | 267.5 | 5091 | 11.48 ± 0.99 | 172.94 ± 2.47 | 2.0 |
| 2021 Oct 08.979 | *BC* | 48.2 | 267.5 | 5091 | 11.48 ± 0.99 | 172.94 ± 2.47 | 2.0 |
| 2021 Oct 08.980 | *RC* | 48.2 | 267.5 | 5091 | 11.23 ± 0.54 | 179.00 ± 1.37 | 2.0 |
| 2022 Feb 6.897 | *r-sdss* | 10.5 | 133.5 | 20,568 | −1.86 ± 0.04 | 153.40 ± 0.83 | 6.0 |
| 2022 Feb 27.781 | *V* | 19.2 | 113.0 | 20,670 | 0.27 ± 0.08 | 146.06 ± 8.57 | 2.0 |
| 2022 Feb 27.781 | *R* | 19.2 | 113.0 | 20,670 | 0.23 ± 0.04 | 83.75 ± 4.48 | 2.0 |
| 2022 Feb 27.808 | *B* | 19.3 | 113.0 | 20,670 | 1.22 ± 0.05 | 89.13 ± 1.22 | 2.0 |
| 2022 Feb 27.810 | *I* | 19.3 | 113.0 | 20,670 | 0.76 ± 0.24 | 75.12 ± 8.86 | 2.0 |

### 4.2. *Aperture polarimetry*

The results of linear-polarimetric observations of comet 67P/C–G during the 2021/22 apparition were compared with data from the PDS Comet Polarimetry Database (Kiselev et al., 2017). Table 2 presents measurements, including the observation time, filter, phase angle, position angle of the scattering plane $\varphi$, diaphragm size in km, measured degree of linear polarization $P$ and its error $\sigma_P$, position angle $\theta$ of the polarization plane in the equatorial coordinate system and its error $\sigma_\theta$, and telescope.

For visualization, we constructed the individual phase-angle dependences of the polarization degree of comet 67P/C–G, including measurements obtained in the previous apparitions: 1982, 2009, 2015, and 2021 (Fig. 5). Data from all four apparitions are indicated with different symbols and colors, as shown in the figure. Some scatter in the data can be attributed to the observational errors (which can reach 1%) and differences in apertures and filters. There are no polarization measurements of comet 67P/C–G in previous apparitions within the range of phase angles approximately 30°–50°, except for our and independent data of Gray et al. (2024). The polarization of the comet measured at phase angles from ~ 28° to ~ 50° during 2021/22 apparition appears systematically higher than the standard (reference) phase-angle curve for high-$P_{\max}$ comets derived by Kiselev et al. (2015) for measurements in the red range (black line in Fig. 5). The plot obtained from the observations in 2021/22 apparition (Fig. 5) includes data from Gray et al. (2024) obtained by the authors during the 2021/22 and 2015/16 apparitions. These data were included to better constrain the polarization behavior at small phase angles and to compare the resulting negative branch of polarization with the fit obtained for high-$P_{\max}$ comets (Kiselev et al., 2015). It can be seen that, during the 2015/16 apparition, the measured polarization of the comet at small phase angles does not differ significantly from the standard phase-angle curve for dusty comets.

The upper-left panel represents measurements from the 1982 apparition: triangles (Myers and Nordsieck, 1984) and diamonds (Chernova et al., 1993). The upper-right panel shows data obtained during the 2009 apparition: inverted triangles (Hadamcik et al., 2010) and squares (Stinson et al., 2016). The lower-left panel presents data obtained during the 2015/16 apparition: polygons (Hadamcik et al., 2016), circles (Rosenbush et al., 2017), and filled triangles (Kwon et al., 2022). The lower-right panel shows the data from Gray et al. (2024), indicated by open red triangles for the 2015/16 and 2021/22 apparitions. Our polarization measurements of the comet through the $B$, $V$, $R$, $I$, $BC$, and $RC$ filters during the 2021/22 apparition are marked with gray circles, whereas the weighted average polarization value for these filters for each date is shown by black circles. Green open circles show our measurements obtained with the 6-m telescope in 2021/22. The solid curve indicates the average phase-angle dependence of polarization for high-$P_{\max}$ comets in the red continuum taken from (Kiselev et al., 2015).

In general, the phase-polarization curve of comet 67P/C–G resembles that of high-polarization, dusty comets (Kiselev et al., 2015), although the available data suggest that our observations from 2021/22, as well as those by Gray et al. (2024), show polarization values exceeding those usually observed for high-polarization comets. The lack of data at large phase angles in previous observations does not allow us to determine whether this is a typical behavior of 67P/C–G or a specific feature of the 2021/22 apparition.

## 5. Numerical modeling of color and polarization variations within the coma

To extract the properties of dust particles in the coma and tail of comet 67P/C–G from the color and polarization data, we model the dust using combined light-scattering and dynamical approaches. For light scattering calculations of small particles with radius $R \leq 5$ μm, we employ the Fast Superposition T-matrix method (FaSTMM; Markkanen and Yuffa, 2017), which solves the Maxwell equations without any physical approximations. Although this numerical solution is asymptotically exact, its computational cost rises rapidly with particle size. Therefore, for particles larger than up to 1 cm, we employ an approximate Monte Carlo radiative transfer approach tailored to discrete random media. This method requires as inputs the incoherent scattering and absorption coefficients, the incoherent scattering phase function of an elementary volume, and the coherent effective refractive index. These quantities are computed with the FaSTMM, and are then fed into the Monte Carlo radiative transfer solver SIRIS4 (Väisänen et al., 2018) as described by Markkanen et al. (2018).

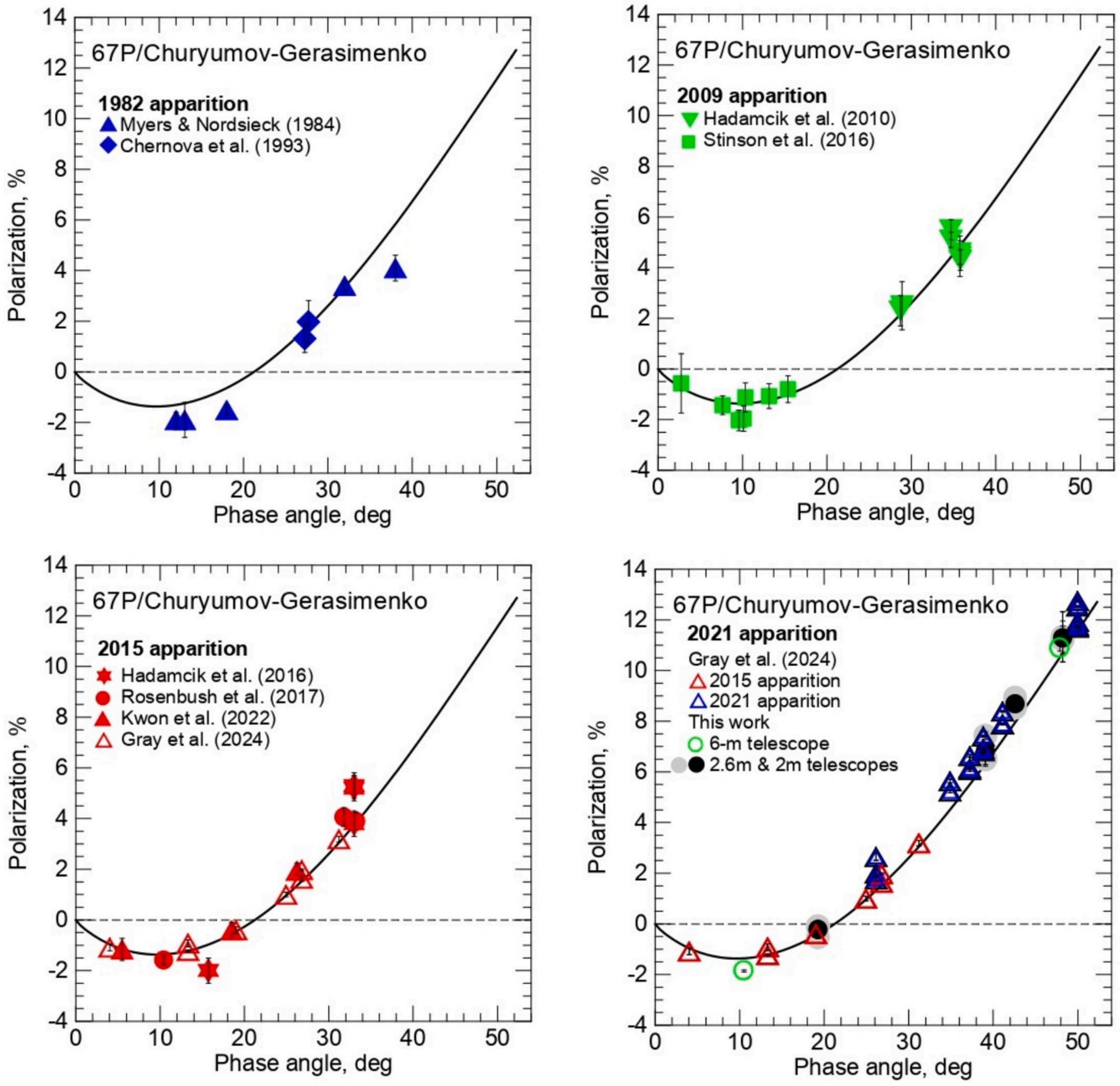

**Fig. 5.** Polarization of comet 67P/Churyumov–Gerasimenko as a function of phase angle for different apparitions.

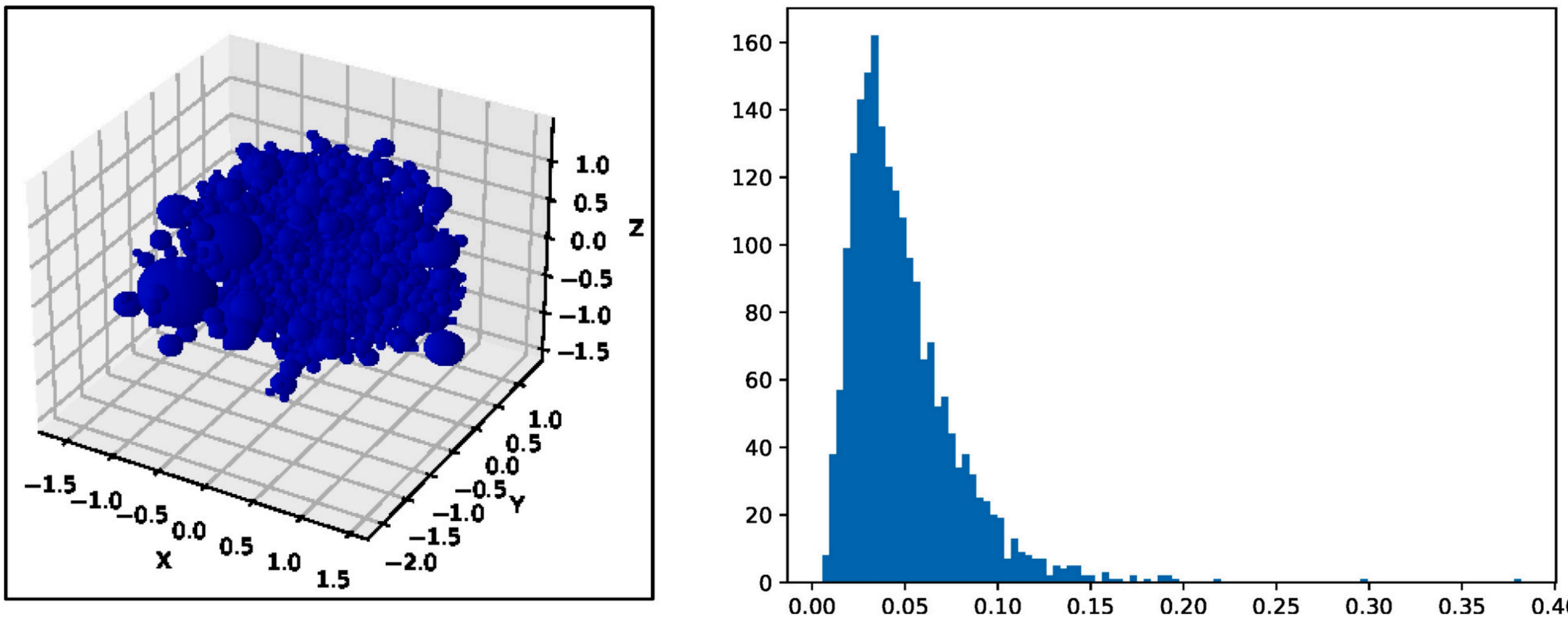

**Fig. 6.** An illustration of a BPCA aggregate made out of 2048 monomers (left) and the histogram of its monomer sizes (right).

### 5.1. Particle model

The light-scattering model represents particles as ballistic particle-cluster aggregates (BPCA) with effective radius $R$ = sqrt(5/3)*gyration radius. Each aggregate consists of spherical monomers whose radii follow a log-normal size distribution with the mean radius of $r = 50$ nm and the standard deviation of $\sigma$=30 nm. The monomer-size distribution is consistent with in-situ measurements of the ROSETTA/MIDAS instrument (Mannel et al., 2019). Fig. 6 illustrates a BPCA aggregate and its monomer-size histogram. The moderately porous BPCA aggregates are selected because they can reproduce the negative polarization branch of the comet 67P/C–G. Highly porous BCCA aggregates would result in much narrower negative polarization branch, while dense aggregates or solid particles would reproduce blue polarimetric color, inconsistent with the 67P/C–G observations.

The monomer composition is assumed to be a mixture of silicate mineral and organic refractory material. For magnesium-iron silicates, the refractive index $m$ typically has a real part around 1.6–1.7 and a small imaginary part between 0.0001 *and* 0.01, depending on the iron content (Dorschner et al., 1995). The refractive index of cometary organic refractories is poorly constrained; literature-reported values range from $m = 1.5 + i0.05$ to $m = 2.4 + i0.8$ (Jenniskens, 1993; Li and Greenberg, 1997). The lower values correspond to weakly processed organics, while higher values indicate more heavily processed material.

### 5.2. Light scattering modeling results

First, we determined a refractive index that allows our simulations to reproduce both the aperture-averaged polarization phase curve and the geometric albedo in red light, that is consistent with the nucleus geometric albedo. We also required the organic-to-silicate mass ratio to be close to the value measured by Rosetta/COSIMA (50% organics and 50% silicates in mass) (Bardyn et al., 2017). For the silicate component, we adopted the refractive index of $Mg_{0.95}$ $Fe_{0.05}$ $SiO_3$. Because silicates absorb far less than organics, the light-scattering properties in visible wavelengths are insensitive to the specific silicate mineral used in the modeling. When using the refractive index of heavily processed organic matter (Jenniskens, 1993), the model yields an excessively high polarization degree and a geometric albedo that is too low. Conversely, fitting the polarization phase curve with weakly processed organic material (Jenniskens, 1993) produces a color that is too red and a geometric albedo that is too high. The refractive index derived by Li and Greenberg (1997) provides the best agreement with both the observed polarization and color, while yielding a geometric albedo is consistent with the nucleus albedo. The resulting refractive using the Maxwell-Garnett mixing rule with 50% of organics and 50% of silicates is $m = 1.754 + i0.157$ for $\lambda = 440$ nm, and $m = 1.778 + i0.133$ for $\lambda = 650$ nm.

Fig. 7 shows the computed polarization phase curves for different particle sizes, together with the observed aperture-averaged polarization degree of 67P/C–G reported by Gray et al., 2024. To investigate spatial structures in the coma, Fig. 8 plots the polarization and color as a function of particle size for the phase angles of 10.5 and 48 degrees. It is evident from Fig. 8 that beyond a certain size threshold, both color and polarization become effectively constant. This behavior arises when the particle size exceeds the photon extinction length scale within the aggregate. Hence, if variations in color and polarization are observed in the coma, they are likely due to particles smaller than 10 μm.

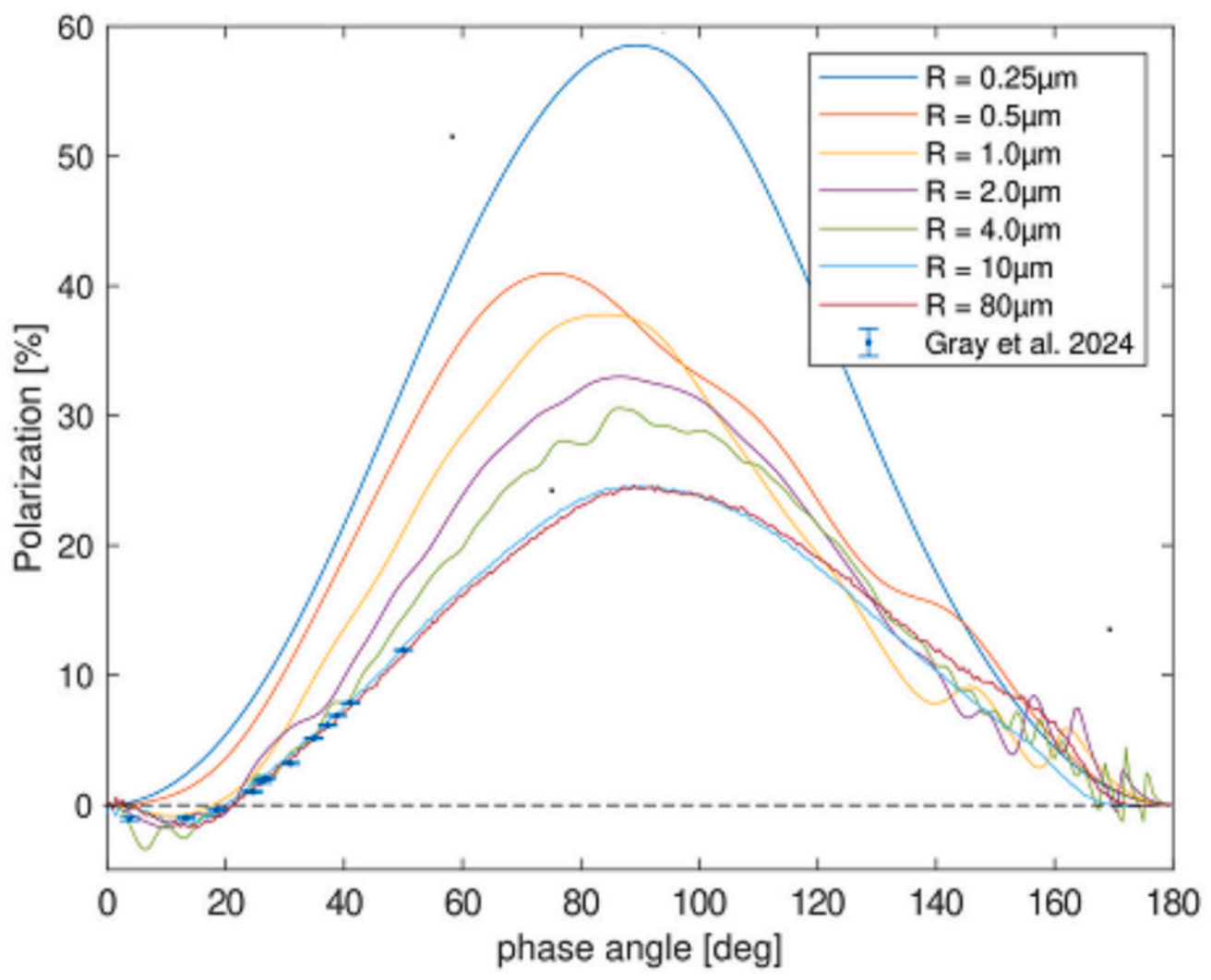

**Fig. 7.** Modeled polarization phase function for different aggregate sizes, including the aperture-averaged polarization measured for comet 67P/Churyumov–Gerasimenko by Gray et al. (2024).

### 5.3. Dust dynamics model

To solve the number density of dust particles within the camera's field of view, we implemented a Monte Carlo dynamical dust model. The model includes the gravitational forces of the Sun and planets, as well as solar radiation pressure acting on the dust particles. In the model, particles are ejected from a spherical surface of radius 10 km centered on the comet's nucleus, with an initial radial velocity

$v_e = \gamma \sqrt{Q_{gas}/r^{\nu}} cos(\theta)$,where the total gas production rate depends on the heliocentric distance $r_h$ as $Q_{gas} \propto r_h^{-6.5}$ as derived from Rosetta measurement by Läuter et al. (2020). Thus, the initial velocities of dust particles derive from a simple gas drag model, while the radius of the emission surface is set to 10 km where gas and dust decouples, and the solar radiation pressure and gravity becomes dominating forces (Agarwal et al., 2024). Also, the emission velocity is assumed to have the cos($\theta$) dependence, in which $\theta$ is the angle from the subsolar point. The parameter $\nu$ is a free parameter that can be related to the size-dependent gas drag coefficient. The directional dependence of the number density of emitted dust is assumed to follow the cosine law, i.e., the number density is highest at the sub-solar point and decreases as a cosine function. No dust emission on the night side of the nucleus is assumed. The heliocentric dependence of the dust production rate is assumed to follow the water gas production rate, and the particle size distribution is assumed to be independent of the heliocentric distance.

We perform a Monte Carlo simulation by releasing a large number of dust particles (around 0.1–1 million particles/day for each particle size) sampled from the prescribed velocity and number density distributions. Particle trajectories are propagated to the observation time using the open-source package Kete (Dahlen et al., 2025). With each particle's position at the time of observations and precomputed light-scattering properties, we synthesize artificial intensity, color, and polarization maps as they would appear to the telescope. To emulate atmospheric seeing and telescope's point spread function effects, we convolve the synthetic intensity images with a two-dimensional Gaussian kernel.

### 5.4. Modeling results

First, we study the observations of 6 October 2021. The dynamical solution is computed for particle size covering the range 0.1 μm – 1 cm with varying $\nu$ and $\gamma$ parameters, describing the initial velocity of the particles. The parameters $\nu = 1.1$ and $\gamma = 0.175$ control the shape of the coma and the dust number density within the coma; hence, they can be relatively easily constrained for a specific dust model. The emitted dust size distribution is described by a power law with three parameters: the minimum size, maximum size, and power-law index. In practice, we compute and store the dynamical solutions for 50 different particle size bins and construct the final image by weighting the number densities in each bin according to the assumed power-law distribution. This approach allows us to efficiently test different size distributions without rerunning the dynamical simulations.

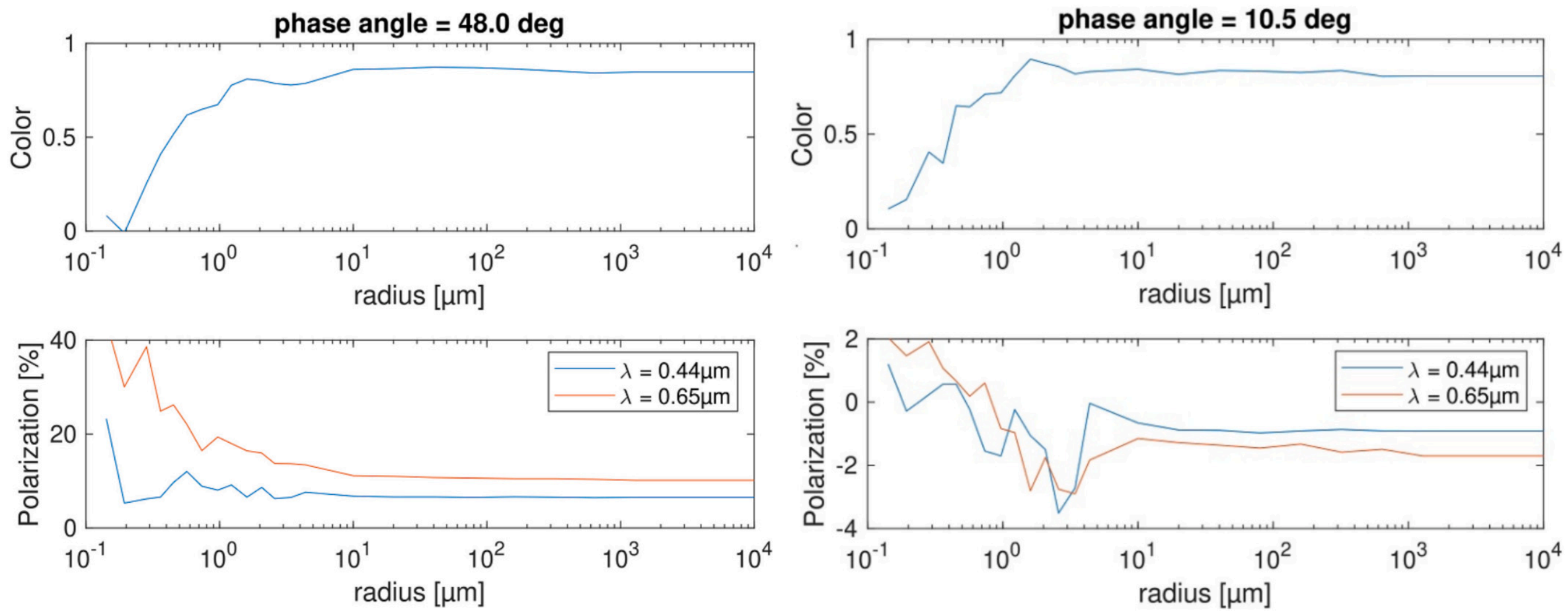

**Fig. 8.** Modeled polarization and color as a function of aggregate size for the phase angles of 10.5° and 48°.

Fig. 9 shows the simulated particle number densities for six representative particle sizes, assuming that continuous activity began 12 months before the observation. It clearly demonstrates the dynamic sorting effect, which produces a spatially varying particle size distribution within the coma (Agarwal et al., 2024). Small particles are rapidly accelerated by gas drag and are strongly affected by radiation pressure, whereas large particles are only weakly influenced by these forces and form a narrow tail of slowly moving material along the negative velocity vector. Importantly, small particles leave the field of view within hours to days, while large particles can remain for months or even years. As a result, the observed coma structure depends strongly on the comet's activity history, including the dust production rate and the initial particle velocities as functions of heliocentric distance.

The observed color and polarization maps (Figs. 1, 3, and 4) show decreasing color and increasing polarization with distance from the nucleus. The dynamical model predicts a decrease in particle size with increasing distance. According to the light-scattering model (Fig. 8), color increases and polarization increases with decreasing particle size. Thus, observations are consistent with a decrease in particle size caused by the dynamic sorting effect. The color index at the outer edge of the coma is approximately 0.2 mag (Fig. 1). Since only the smallest particles reach these regions in the dynamical model, this implies a minimum particle size of about $r_{min} = 0.25$ μm, consistent with the color index in Fig. 8. Adopting $r_{min} = 0.25$ μm, a power-law index of $-$ 3.3 provides the best fit to the aperture-averaged polarization and color. The maximum particle size has little effect on the polarization and color, but can be constrained to $r_{\max} = 2.6$ mm by comparison with the observed tail profile. Larger values of $r_{\max}$ produce a narrow, bright feature along the negative velocity vector, whereas smaller values result in a broader and fainter tail.

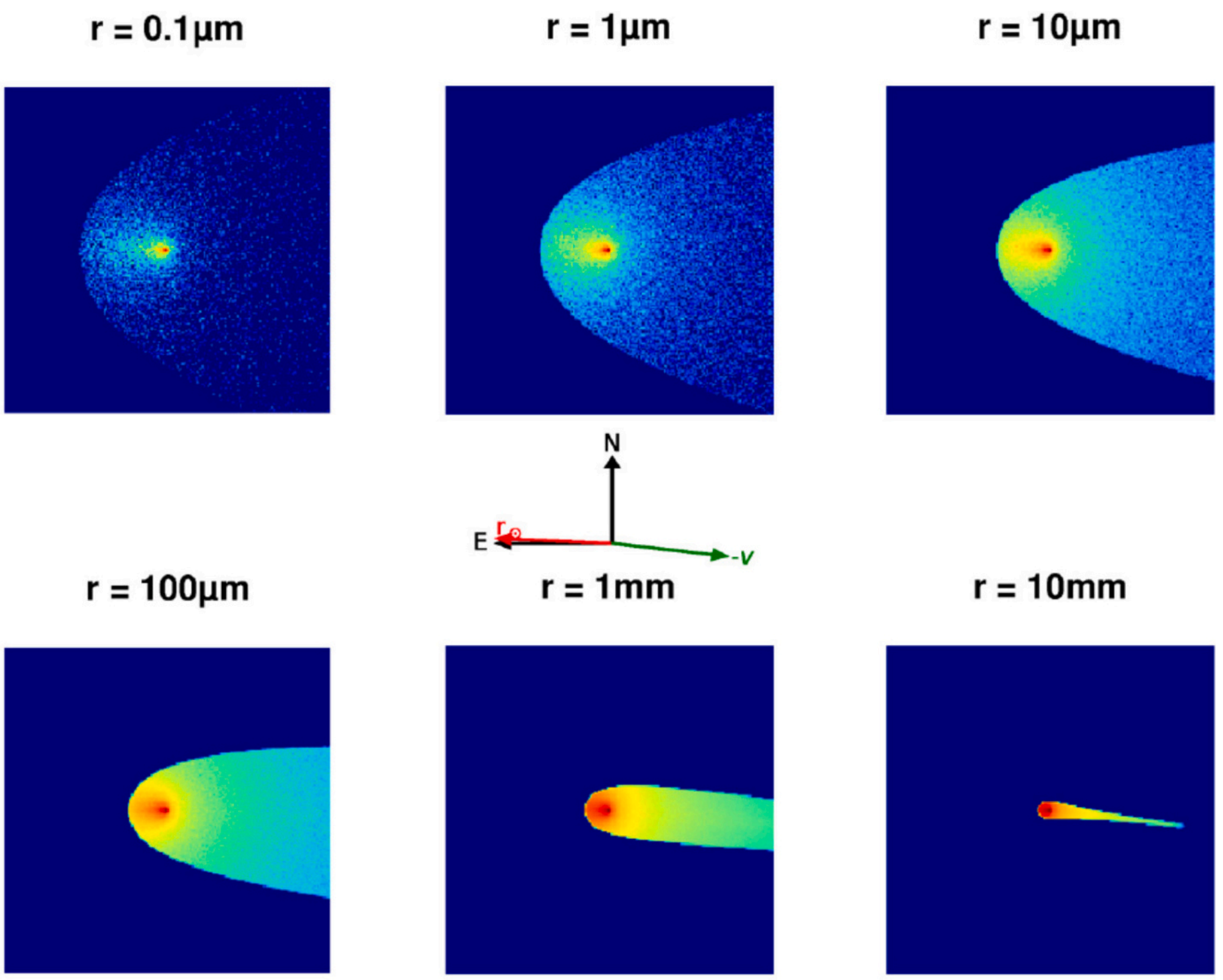

**Fig. 9.** Number densities of particles of different sizes simulated for 6 October 2021. The relative contribution of each size bin to the total number density depends on the assumed size distribution of emitted particles. From smallest to largest, the particles in the field of view are emitted no later than 1 day, 3 days, 2 weeks, 1 month, 3 months, and 12 months prior to the observation epoch, respectively.

Fig. 10 shows the simulated color and polarization maps for 6 October 2021. Overall, they reproduce the main large-scale features of the observations: color decreases from the nucleus toward the outer coma, while polarization shows a moderate increase. The model predicts high polarization at the outer edge of the coma; however, this feature is not evident in the observations, likely due to low signal-to-noise ratios. These results indicate that the large-scale coma structure is primarily governed by the dynamic sorting effect, driven by gas drag, solar radiation pressure, and gravity. However, the model overestimates the rate of increase in polarization and underestimates the rate of decrease in color with distance from the nucleus.

This discrepancy arises from a strong correlation between polarization and color in the light-scattering model. Modifying the aggregate porosity could improve the model, but this would require development of a new aggregation algorithm with controlled porosity.

An alternative explanation is the sublimation of organic material in small particles. Such particles can become superheated, leading to enhanced sublimation rates compared to larger particles. The loss of organics would reduce the imaginary part of the refractive index, thereby decreasing both polarization and color produced by small particles. A more detailed assessment of this process requires extensive modeling and new observations of distributed gas sources in the coma.

Note that the model cannot reproduce small-scale coma structures, such as jets, due to simplifying assumptions including a spherical, homogeneous nucleus and an idealized outgassing model. As a result, small-scale variations near the nucleus observed in polarization maps remain unexplained.

The observations from 6 February 2022 cannot be reproduced using the same model as for 6 October 2021 (compare Figs. 9 and 11). Larger particles (up to ~1 cm) are required to match the brightness profile, particularly the narrow structure along the projected negative velocity vector and the absence of an anti-solar tail (see Fig. 4 in Paper I). Since the comet was receding from perihelion at that time, the coma is unlikely to be dominated by small micrometer-sized particles, as evidenced by the lack of an anti-solar tail. Instead, it is natural that the scattering cross section is dominated by slowly moving large particles released near perihelion, when the comet was more active.

The low negative polarization (~ $-$ 1%) observed near the nucleus and in the narrow tail structure (Figs. 3 and 4) supports the presence of large particles (Figs. 7 and 8). However, the high negative polarization in the outer coma remains difficult to explain. Such polarization typically requires particles in the ~1–10 μm size range (Fig. 8), yet particles of this size should be strongly affected by radiation pressure and form an anti-solar tail directed toward the southeast, which is not observed. This discrepancy highlights the importance of combining the dynamical and light-scattering models to avoid misinterpretation. One possible explanation is the presence of less absorbing particles, for which the polarization minimum shifts to larger sizes (~10–100 μm), allowing the observations to be reconciled with the dynamical constraints. Unfortunately, the absence of simultaneous color observations on 6 February 2022 limits our ability to further constrain the particle properties.

## 6. Discussion

### *6.1. Key patterns in the color and polarization of comet 67P/C–G during its apparitions in 2021/22 and 2015/16*

A comparative analysis of the color and polarimetric properties of comet 67P/C–G during its 2015/16 and 2021/22 apparitions reveals both stable dust characteristics and significant changes associated with different levels of nucleus activity. In both periods, the same global trend is observed: the coma becomes bluer with increasing distance from the nucleus. The dynamic sorting (dust acceleration by gas drag near the nucleus and combined effect of solar gravity, and radiation pressure) is most significant in producing a growing contribution of fine dust in the outer regions. Fragmentation may also play some role, at least during the 2021/22 apparition, although stronger fragmentation results in simulated polarization and color maps that are inconsistent with the observations. In the tailward direction, the dust remains significantly redder in both apparitions, indicating the dominance of larger, dynamically older particles. Thus, a qualitative comparison of the color maps suggests that the fundamental mechanisms shaping the coma structure remained broadly unchanged between the two returns of the comet.

At the same time, quantitative differences between the two perihelion passages are substantial. In the central coma, the color values ($g$–$r$) in 2021/22 and 2015/16 are similar, about ~0.8 mag, indicating comparable optical properties of dust near the nucleus. However, the radial color gradients and spatial structures differ. In 2015/16, ($g$–$r$) decreased to ~0.4 mag in the outer coma, and the color maps revealed pronounced local inhomogeneities, asymmetries, and jets. In contrast, during 2021/22, the bluing of the coma occurred more uniformly and on smaller spatial scales: within the jets and central coma, (*g*–*r*) decreased to ~0.2 mag already at distances of ~17,000–19,000 km from the nucleus. The tail color, meanwhile, remained consistently red in both periods, with (*g*–*r*) ≈ 0.8 mag.

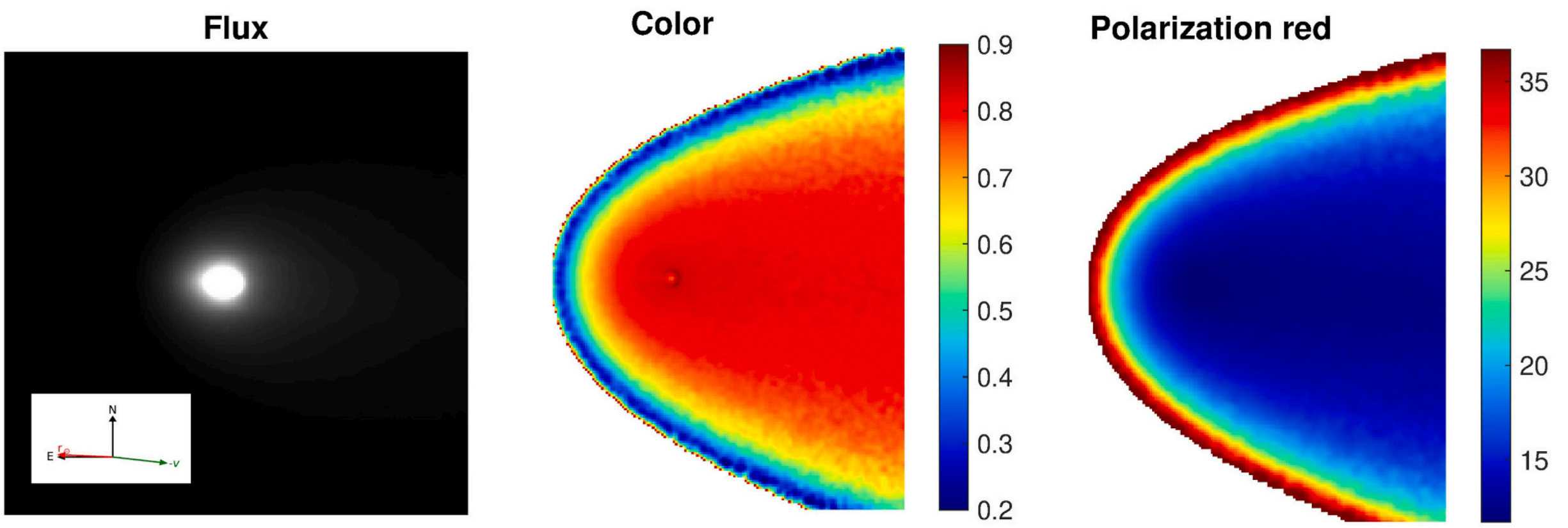

**Fig. 10.** Simulated color and polarization maps of comet 67P/Churyumov–Gerasimenko on 6 October 2021, as well as extracted slices along and perpendicular to the negative velocity vector of the nucleus.

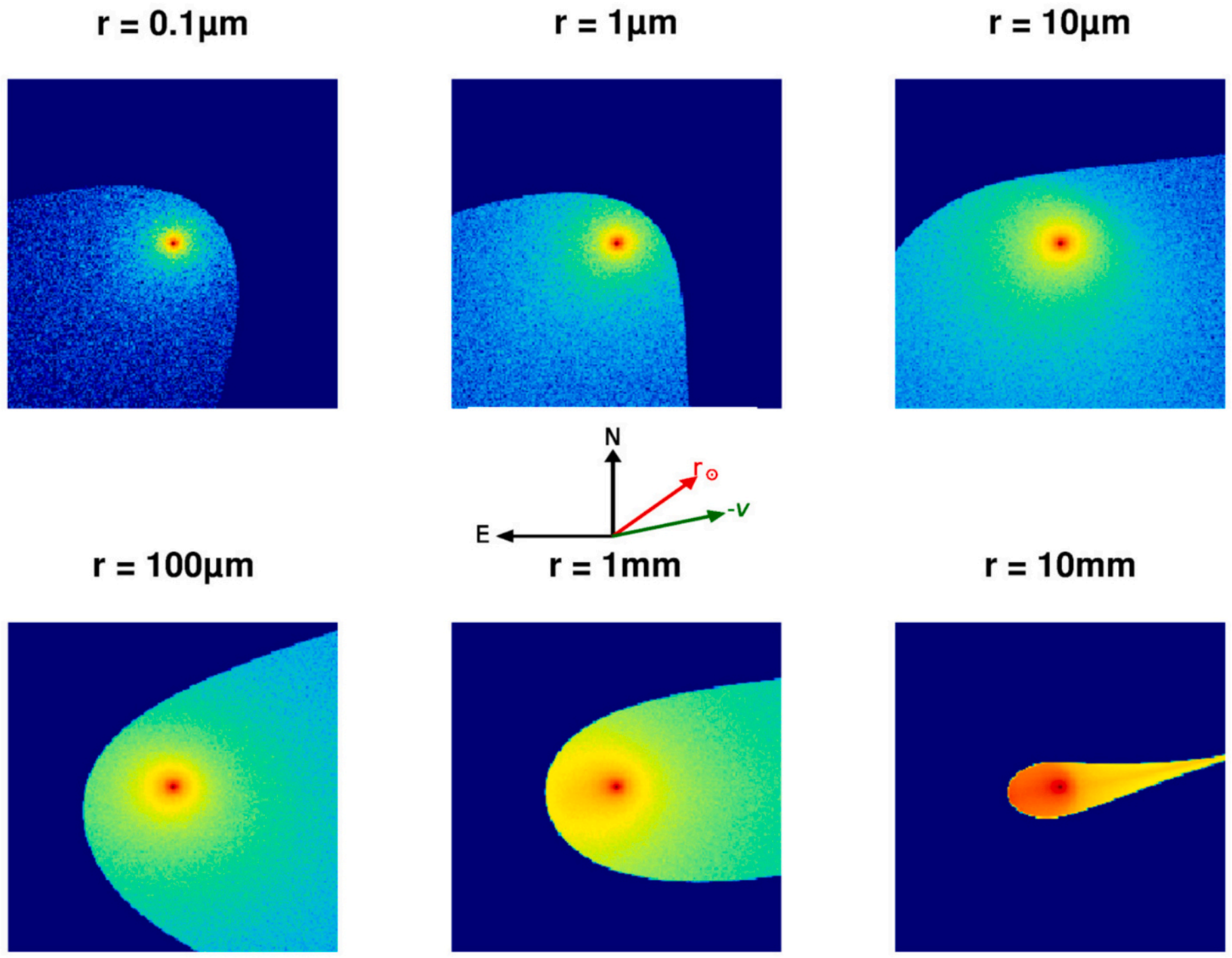

**Fig. 11.** Same as in Fig. 9, but for 6 February 2022.

The polarimetric data further support these conclusions, indicating that the dust behavior differed between the two apparitions. At phase angles of about 31–33°, observations in 2015 showed a complex and spatially heterogeneous polarization structure, with sharp variations in polarization degree across different coma regions. In the central region, the linear polarization reached ~8%, then decreased to ~2% at distances of several thousand kilometers, and again increased to ~8% in the outer coma. In contrast, in 2021, at a larger phase angle ($\alpha \approx 47.9°$), the polarization was high and nearly uniform across the coma and tail, with typical values of ~10–12% and only minor deviations in the jet. Such smoothing of the polarization field indicates a more homogeneous distribution of dust properties in the coma.

Similar differences are observed at small phase angles. After perihelion in 2016, the negative polarization in the coma was deep and structured, reaching −3 to −4% in the outer regions. In 2022, at a phase angle of about 10.5°, the negative polarization was generally weaker and smoother: along the main coma structures, its magnitude varied mostly between −1 and − 2% and increased in absolute value much more slowly with distance from the nucleus. At the same time, within the coma, it decreased rather sharply to −1 to −2%, indicating differences in dust properties. A similar spatial pattern of polarization during the 2021/22 apparition was also reported by Gray et al. (2024) for phase angles of 26.2–50°, confirming differences in the comet's polarimetric behavior between the two apparitions of comet 67P/C-G.

In summary, the general dust properties persist between the two apparitions: a red inner coma, progressive bluing with distance from the nucleus, a consistently red tail, and contrasts between jets and the background coma. At the same time, in 2021/22 these structures are significantly weaker: the coma is more uniform in color, the polarization field is smoother, and the negative polarization is more pronounced. According to Fig. 5, comet 67P/C–G exhibits higher positive polarization compared to the average phase-angle dependence for high-*P*max comets in the red continuum (Kiselev et al., 2015). Together with differences in dust production, specifically $Af(0°)\rho$ values in the *r*-sdss filter of about 500 cm in 2015 and higher values of about 740–940 cm in 2021 (see Table 5 in Paper I), these characteristics indicate a noticeable change in nucleus activity and dust properties between the two perihelion passages.

### 6.2. *Features of the polarimetric behavior of the comet*

The most striking peculiarity of the 2021/22 apparition is the difference in the comet's polarimetric behavior, namely that the cometary activity did not produce pronounced polarimetric inhomogeneities. It is quite common that cometary activity visible in brightness, for example in the form of jets, appears differently in polarization (e.g., Rosenbush et al., 2017). This is because photometric features often reflect variations in the amount of dust, i.e., enhanced dust abundance, whereas polarization features indicate differences in the physical properties of the dust, such as particle size, composition, or structure. The presence of photometric jets in both apparitions, combined with the absence of significant polarization inhomogeneities in 2021/22, suggests that although certain regions of the nucleus remained active, they produced dust with properties similar to those of the surrounding coma. In contrast, during the 2015/16 apparition, some active regions released dust with distinct physical characteristics, and these regions appear to have ceased activity by 2021/22. Whether this reflects a peculiarity of the 2021/22 return or a more general long-term trend of comet 67P/C–G can only be clarified by similar observations during future apparitions.

Another important result of the 2021/22 observations is that the positive polarization reliably exceeds the values typically observed for high-polarization comets. The data obtained during this apparition are the first at sufficiently large phase angles for this difference to become clearly evident. This raises the question of whether the higher-than-average polarization is an intrinsic property of comet 67P/C–G or a specific feature of the 2021/22 apparition.

Traditionally, high polarization has been associated with the dominance of small dust particles, as in the case of comet C/1995 O1 (Hale–Bopp). However, as discussed by Kolokolova et al. (2007), such comets also exhibit a strong silicate feature in the mid-infrared. In contrast, the mid-infrared spectrum of comet 67P/C–G shows a weak silicate feature (Kelley et al., 2006). In this context, the high polarization of the cometary dust may result from an enhanced abundance of heavily processed organic material in the dust composition (e.g., Levasseur-Regourd et al., 2018; Kolokolova et al., 2024). Although a high content of processed organics does not contradict the modeling results presented in Section 5. Small particles (such as those in comet Hale-Bopp) would produce a bluer coma and would form a bright anti-solar tail. We therefore rule out a predominance of small particles. However, in the absence of mid-infrared observations for this apparition, firm conclusions remain limited. Future polarimetric observations of comet 67P/C–G, particularly at large phase angles and in combination with mid-infrared spectroscopy, will be essential to better constrain its dust properties.

### 6.3. *Spatial distribution of color in the coma*

The coma exhibits a significantly increased color index in the central region near the nucleus, reaching values up to almost 0.9 mag (Fig. 1). It is important to note that the observed color represents an average of the dust properties along the line of sight. Since the color index decreases to approximately 0.2 mag at the edge of the coma, the intrinsic color index of the innermost coma should be higher and likely exceeds 1.0 mag. This is an interesting result because direct measurements of the nucleus surface properties (Fornasier et al., 2015) show that at a phase angle of 47.9°, the spectral slope of the nucleus reflectivity is 16.3%/1000 Å, corresponding to a color index of 0.26 mag. As the coma is observed in reflected sunlight, the solar color index of 0.46 mag in the *g*-sdss and *r*-sdss filters (Willmer, 2018) should be added, yielding a nucleus color index of 0.72 mag. This may indicate the presence of an excess of large particles near the nucleus, allowing them to remain in the inner coma for a long time. When near the nucleus, these particles lose a significant fraction of their volatiles and therefore become redder in color. As a result, the inner coma appears redder than the surface of the nucleus.

The color map obtained on 6 October 2021 shows that the maximum color index is located near the nucleus (Figs. 1 and 2). Most likely, this indicates that freshly ejected dust is dominated by large red particles. The profile along the tail direction does not show a noticeable decrease in dust redness, suggesting that the particles remain predominantly larger than 10 μm (see Fig. 8). Such particles are only weakly affected by solar radiation pressure, allowing them to remain close to the comet's orbit for extended periods and to form a trail. The difference between the 2015 and 2021 profiles may be explained by the different heliocentric distances at the two epochs: 1.615 au on 8 November 2015 versus 1.254 au on 6 October 2021. The smaller heliocentric distance in 2021 could have led to more intense outgassing and the release of a larger number of large particles, which subsequently contributed to the formation of the neckline structure observed on 6 February 2022, and described in Paper I. However, there is some effect of the much larger difference in geocentric distance. On 8 November 2015, the comet was at a geocentric distance of $\Delta = 1.80$ au, while on 6 October 2021, at $\Delta = 0.48$ au. The 2021/22 observations had better spatial resolution and therefore allowed us to better resolve the inner coma area (see Paper I) that could influence the observed dust distribution and brightness.

A comparison of the jet activity on 6 October 2021 and 8 November 2015 (Rosenbush et al., 2017) reveals significant differences. In 2021, the jet is not visible on the color map against the surrounding coma, even after applying digital filters, whereas in 2015, the jets were clearly identified after applying a rotational-gradient filter. The 2015 jets also differ in the rate at which their color index decreases with distance compared to the surrounding coma. These distinctions indicate differences in the dust properties within the jets during the two apparitions. Based on the jet modeling results (Paper I), this may be explained by different source regions on the nucleus surface (see Paper I: Fig. 7a for the 2021 jet and Fig. 16 for the 2015 jets). This suggests significant compositional heterogeneity of the regions where the jet sources were located during the two returns.

Due to the orientation of the rotation axis, the northern and southern hemispheres experience markedly different insolation. Near perihelion (2 November 2021), the southern hemisphere was predominantly illuminated (Jorda et al., 2016; Keller et al., 2017). During our observations on 6 October 2021, the subsolar latitude was −38°, indicating that the southern hemisphere received most of the solar radiation. This implies prolonged and more intense heating of the southern hemisphere. Higher temperatures should lead to increased gas pressure, making it possible to eject smaller (i.e., optically bluer) grains, which are typically more difficult to lift due to their higher tensile strength (see Groussin et al. (2019) and Attree et al. (2025) for a discussion of the size dependence of tensile strength). This is why Fig. 2 (b and c) shows that the dust in the southern hemisphere is bluer than in the northern hemisphere.

The ejection of redder dust from the northern hemisphere was also detected by the Rosetta mission using the OSIRIS WAC camera on 22–23 August 2015 (Kolokolova et al., 2024; Kleshchonok et al., 2025). At that time, the subsolar latitude was about −50°, meaning the southern hemisphere was predominantly illuminated. Nevertheless, the dust coma on the non-illuminated side of the nucleus, largely corresponding to the northern hemisphere, exhibited a noticeably redder color. The difference in color index between the illuminated and non-illuminated sides was small, on the order of 0.05–0.1 mag in the *UV375* and *VIS610* filters. This indicates that the color asymmetry originates in the immediate vicinity of the nucleus, within the inner coma, and then extends outward.

Finally, the bluer regions in the southern hemisphere correlate with lower polarization (cf. Fig. 2b and Fig. 3a), indicating a correlation between color and polarization. As shown in previous studies (e.g., Kolokolova et al., 2001), a positive correlation between color and polarization suggests compositional differences, whereas variations in particle size typically lead to an anticorrelation between these parameters. The observed correlation in the southern hemisphere may therefore indicate that this hemisphere produced less processed (less absorbing; see Section 5.4) dust particles. This may occur because strong illumination leads to increased erosion of the surface, exposing fresh subsurface material.

## 7. Conclusion

We present and discuss photometric and polarimetric data acquired during the 2021/22 apparition of the Jupiter-family comet 67P/Churyumov–Gerasimenko. The observations were obtained at a phase angle of 47.9° (6 October 2021, before perihelion) and at a small phase angle of 10.5° (6 February 2022, after perihelion), enabling us to investigate both positive and negative polarization. Photometric data in the *g*-sdss and *r*-sdss filters, together with polarimetric data in the *R* filter, were used to construct color and polarization maps, which were subsequently analyzed using light-scattering and dynamical modeling. To derive the phase-angle curve of polarization, we carried out photoelectric measurements of the polarization of comet 67P/C–G using different apertures and filters within the range of phase angles of 19.2°–48.2°. Analysis of these observations, combined with numerical simulations, allows us to draw the following conclusions:

- The color image appears relatively homogeneous, exhibiting only a smooth decrease in the color of the coma and jet from approximately 0.85 mag near the nucleus to about 0.2 mag at cometocentric distances of 20,000 km on the first observing date. At the same time, the color of the tail remains nearly constant (∼0.75 mag) from the nucleus out to a cometocentric distance of ∼35,000 km.
- By applying dedicated digital filtering techniques, a color asymmetry was identified in the 2021 coma color map. This asymmetry is interpreted as a result of the differing insolation of the northern and southern hemispheres of the nucleus, resulting in the lifting of smaller, and, due to higher erosion rate, less-processed subsurface particles from the more strongly illuminated southern hemisphere.

- The polarization maps likewise do not display any significant azimuthal structures. Instead, polarization changes smoothly with distance from the nucleus, ranging from about 10–12% on 6 October 2021 and from about −1% to −2% on 6 February 2022, except in where the negative polarization reaches values of up to −4%.
- The colorimetric and polarimetric data were modeled by representing dust particles as moderately porous BPCA-type aggregates of various sizes. The best agreement with the observed color, polarization, and albedo was achieved for aggregates composed of equal fractions (50%) of silicates and processed organics. Light-scattering calculations were performed using the FaSTMM technique and subsequently employed as input for the Monte Carlo radiative transfer solver SIRIS4. Monte Carlo methods were also applied in dynamical modeling to describe dust motion within the coma.
- The modeling results indicate that the coma is dominated by relatively large particles with sizes exceeding 10 μm. This lower size limit is supported by the finding that both polarization and color become nearly independent of particle size for sizes larger than ~10 μm.
- Small-scale variations detected in the polarization maps are most likely due to locally enhanced abundances of smaller particles, specifically those in the 1–10 μm size range, or to particles containing less absorbing material. The latter explanation appears more plausible, as the color maps reveal bluer regions that approximately coincide with areas of lower polarization.
- With increasing distance from the nucleus, the effective particle size decreases, leading to gradual spectral bluing and an increase in polarization. Interestingly, no comparable changes in color or polarization are observed in the tailward direction. According to our modeling, this behavior can be explained by a dynamic sorting effect: small particles are accelerated to higher speeds than large ones by gas drag, and then they move toward the tail by solar gravity and radiation pressure. Although their sizes decrease with distance from the nucleus, they remain larger than 10 μm.
- We combined our results with aperture polarimetric data obtained during several apparitions of comet 67P/C–G, including those in 2021/22, 2015/16, 2008, and 1982. The resulting composite polarization phase-angle curve shows that the 2021/22 data exhibit polarization values exceeding the typical range observed for high-polarization comets. Since this enhancement is apparent only during the 2021/22 apparition, it remains uncertain whether it is specific to that return or reflects a more general peculiarity of comet 67P/C–G.
- An additional noteworthy outcome of this study is the difference in the polarimetric properties of comet 67P/C–G during its 2021/22 apparition compared to earlier returns. In particular, unlike previous apparitions, both color and polarization display a relatively homogeneous coma, without jets or other distinct features. It is unclear whether this behavior is unique to the 2021/22 apparition or indicative of a longer-term evolutionary trend that may persist in future returns of comet 67P/C–G.

In summary, comet 67P/C–G exhibited smooth radial variations in color and polarization during its 2021/22 apparition, and no pronounced jets or other distinct structures, previously observed in earlier returns, were detected. Numerical modeling suggests that the coma is dominated by large, moderately porous dust aggregates composed of silicates and processed organics, while minor local variations are likely attributable to small admixtures of finer particles or particles containing less processed organics. A comparison with polarimetric data from earlier apparitions indicates that the polarization measured in 2021/22 is unusually high, exceeding values typical of high-polarization comets. These findings imply that the dust properties during this apparition might differ from those observed previously; however, only future multiwavelength photometric and polarimetric observations at subsequent apparitions will clarify whether this behavior is specific to the 2021/22 return or represents a long-term evolutionary trend in comet 67P/C–G.

## CRediT authorship contribution statement

**Vera Rosenbush:** Writing – review & editing, Writing – original draft, Visualization, Supervision, Methodology, Formal analysis, Conceptualization. **Oleksandra Ivanova:** Writing – review & editing, Software, Methodology, Formal analysis, Data curation, Conceptualization. **Johannes Markkanen:** Methodology, Formal analysis, Conceptualization. **Igor Lukyanyk:** Writing – review & editing, Visualization, Software, Investigation, Formal analysis, Conceptualization. **Valerii Kleshchonok:** Writing – review & editing, Visualization, Software, Methodology, Formal analysis. **Ludmilla Kolokolova:** Formal analysis, Conceptualization. **Colin Snodgrass:** Writing – review & editing, Supervision, Methodology, Data curation, Conceptualization. **Elena Shablovinskaya:** Data curation.

## Declaration of competing interest

The authors declare that they have no known competing financial interests or personal relationships that could have appeared to influence the work reported in this paper.

## Acknowledgments

The authors are deeply grateful to the anonymous reviewers, who provided many useful and constructive comments and recommendations, which improved our paper. This work is supported by the Partnership Fund of the University of Edinburgh and Taras Shevchenko National University of Kyiv. The work by OI is supported by the Slovak Grant Agency for Science VEGA (Grant No. 2/0067/26) and by the Slovak Research and Development Agency under Contract No. APVV-24-0076. The research by VR, IL, and VK is supported by Project No. 0126 U002694, and by VR and IL is supported by Project No. 0125 U002319 of the Ministry of Education and Science of Ukraine. JM is financially supported by Deutsche Forschungsgemeinschaft (DFG) Project No: 517146316. The research by VK funded by the Deutsche Forschungsgemeinschaft (DFG, German Research Foundation) – KL 3895/1-1. JM would like to thank Dar Dahlen for his help with the KETE software package. The observations at the 6-m BTA telescope were carried out by E. Shablovinskaya, and at the 2.6-m and the 2-m telescopes were performed by N. Kiselev. We thank them for their assistance in conducting observations and providing results.

## Data availability

Data will be made available on request.